\documentclass[aps,pra,showpacs,twocolumn]{revtex4}
\usepackage{bm, amsmath, amsfonts}
\usepackage{graphicx}
\usepackage{color}

\begin{document}

\title{Determinant representations of spin-operator matrix elements in the XX spin chain and their applications}

\author{Ning Wu}
\email{wun1985@gmail.com}
\affiliation{Center for Quantum Technology Research, School of Physics, Beijing Institute of Technology, Beijing 100081, China}

\begin{abstract}
For the one-dimensional spin-1/2 XX model with either periodic or open boundary conditions, it is shown by using a fermionic approach that the matrix element of the spin operator $S^-_j$  ($S^-_{j}S^+_{j'}$) between two eigenstates with numbers of excitations $n$ and $n+1$ ($n$ and $n$) can be expressed as the determinant of an appropriate $(n+1)\times (n+1)$ matrix whose entries involve the coefficients of the canonical transformations diagonalizing the model. In the special case of a homogeneous periodic XX chain, the matrix element of $S^-_j$ reduces to a variant of the Cauchy determinant that can be evaluated analytically to yield a factorized expression. The obtained compact representations of these matrix elements are then applied to two physical scenarios: (i) Nonlinear optical response of molecular aggregates, for which the determinant representation of the transition dipole matrix elements between eigenstates provides a convenient way to calculate the third-order nonlinear responses for aggregates from small to large sizes compared with the optical wavelength; and (ii) real-time dynamics of an interacting Dicke model consisting of a single bosonic mode coupled to a one-dimensional XX spin bath. In this setup, full quantum calculation up to $N\leq 16$ spins for vanishing intrabath coupling shows that the decay of the reduced bosonic occupation number approaches a finite plateau value (in the long-time limit) that depends on the ratio between the number of excitations and the total number of spins. 
Our results can find useful applications in various ``system-bath" systems, with the system part inhomogeneously coupled to an interacting XX chain.
\end{abstract}


\maketitle

\section{Introduction}
\par The study of quantum spin chains has a long history dating back to Bethe's exact solution of the one-dimensional Heisenberg model in the early 1930s~\cite{Bethe}. With the intention of finding system that bears reasonably close resemblance to the Heisenberg model, Lieb, Schultz, and Mattis~\cite{LSM} introduced the one-dimensional XY model and solved it exactly using the Jordan-Wigner transformation, which changes spin operators into fermions. Later, the authors of Ref.~\cite{LSM} applied a similar second-quantization formalism for fermions to the exact solution of the two-dimensional Ising model~\cite{2DIsing}. Over the past several decades, the fermionic approach has found wide applications in dealing with many-body systems of spin degrees of freedom. Recently, Iorgov \emph{et al.} derived a factorized formula for spin-operator matrix elements between general eigenstates of the transverse Ising model by using the fermionic technique~\cite{Iorgov}.
\par In this work, we will consider a simpler but frequently used quantity, i.e., the spin-operator matrix element (SOME) between two eigenstates of the XX spin chain, which can be calculated by using similar  techniques to that in Ref.~\cite{Iorgov}. In spite of its simple form, the XX model not only models the physics of spins arranged in a row, but it can also describe many other quantum phenomena, such as the repulsive Bose-Hubbard model in the strong interaction limit~\cite{BHM}, quantum state transfer~\cite{Stolze}, coherent excitation transfer in light-harvesting~\cite{fleming} and Rydberg systems~\cite{Rydberg}, and the dynamics of molecular aggregates~\cite{Spano1991,Hanamura}, among others. These SOMEs naturally emerge in a generic class of composite systems consisting of an XX chain (the ``spin bath") with each spin coupled to the other common quantum system (the ``ystem"), e.g., a single bosonic mode~\cite{Hanamura} or a central spin~\cite{PRA2014,PRB2016}. Note that the number of excitations (e.g., the number of spins pointing upward) of the XX chain is conserved. We will focus on two types of SOMEs, both of which are relevant in the study of static or dynamical properties of these hybrid system-spin-bath systems.  The first operator we consider is the spin-lowering operator $S^-_j$ of spin-$j$, which connects an eigenstate with $(n+1)$ excitations to the other eigenstate with $n$ excitations. The other type of operator we are interested in is the product of the spin-lowering and spin-raising operators on two (not necessarily different) sites $j$ and $j'$, namely $S^-_jS^+_{j'}$, which does not induce spin-flip and connects two eigenstates with the same number of excitations. These kinds of matrix elements might first appear in the study of nonlinear response~\cite{Spano1991} and superradiance~\cite{Hanamura} of one-dimensional molecular aggregates, where the Frenkel-exciton model description of the aggregates resembles an XX chain. Recently, Wu \emph{et al}.~\cite{PRB2016} studied the decoherence dynamics of a single qubit in an extended Gaudin model with an interacting spin bath modeled by the XX ring, where linear combinations of the matrix elements of $S^-_j$ emerge as coefficients in the equation of motion of the system.
\par Though the aforementioned matrix elements are important in various physical systems involving the XX chain, their evaluation is not straightforward since the eigenstates of the XX chain generally do not admit simple forms in the spin configuration space, but rather they are filled by the Jordan-Wigner fermions. For example, in the context of a one-dimensional molecular aggregate, the matrix elements of the excitonic transition dipole moment, which is proportional to $\sum_j(e^{i\mathbf{k}\cdot \mathbf{r}_j}S^+_j+\mathrm{H.c.})$ (with $\mathbf{k}$ the wave vector of the light field and $\mathbf{r}_j$ the position of the $j$th monomer, respectively), was shown~\cite{Spano1991} to have a form similar to Eq.~(\ref{f_SS}) (see below) that involves a multi-summation over all configurations of the spatial indices with a fixed number. Meanwhile, it was recognized that these sums are usually difficult to evaluate~\cite{Spano1991}. When the spatial dimension of the aggregate is much smaller than the optical wavelength, so that the phase factors in the transition dipole moment do not depend on the molecular position $\mathbf{r}_j$, the authors of Ref.~\cite{Hanamura} have been able to evaluate the sums through  ``some tedious calculation". Actually, they obtained a factorized formula [see Eq.~(\ref{DI}) below] for the matrix element of the collective spin operator $S^-=\sum_j S^-_j$, which was recently used in the study of enhanced photon capture in ringlike optical emitter systems~\cite{ratchet2015}. There were some attempts to derive this factorized expression through evaluating the multi-sums by using properties of matrix determinants~\cite{math}; however, to the best our knowledge, an explicit proof of Eq.~(\ref{DI}) is still absent from the literature.
\par In this work, we employ a similar fermionic technique developed in Ref.~\cite{Iorgov} to derive the matrix elements of both $S^-_j$ and $S^-_jS^+_{j'}$ between two relevant eigenstates of an inhomogeneous XX chain with either periodic or open boundary conditions. We show that both of them can be expressed as the determinant of some $(n+1)\times (n+1)$ matrix whose entries involve the coefficients of the canonical transformations diagonalizing the model in its Jordan-Wigner fermion representation. For the special case of the homogeneous periodic XX chain, the determinant representation of the matrix elements of $S^-_j$ turns out to be a variant the Cauchy determinant that can be evaluated analytically, and hence leads to the factorized formula discovered in Ref.~\cite{Hanamura}.
\par We next apply the obtained results to two physical problems, namely the nonlinear response of molecular aggregates and the real-time dynamics of an interacting inhomogeneous Dicke model consisting of a single bosonic mode coupled to an XX chain. In the former case, the determinant representations of the SOMEs provide a convenient way to calculate the aggregate transition dipole matrix elements, which are essential for obtaining the nonlinear optical response functions. The advantages of the present method become more apparent when either the aggregate sizes are large compared with the optical wavelength, or higher-order nonlinear responses are considered, for which the transition dipole matrix elements do not admit closed forms anymore and the determinant representations offer an almost unique tool for efficient evaluation of these matrix elements. In the latter case, the proposed hybrid model can properly describe a linear molecular aggregate located in a single-mode cavity~\cite{2016LPP}. In the absence of the nearest-neighbor coupling within the chain, the model reduces to the ordinary inhomogeneous Dicke model that has been studied thoroughly in quantum optical systems~\cite{Loss2009,Loss2010,PRA2009,Fari2012}. By writing the spin-boson interaction in the eigenbasis of the XX chain and the free boson using the obtained SOMEs, we perform full quantum calculation of the system dynamics. In the noninteracting limit with inhomogeneous excitonic energies and a uniform spin-boson coupling, we find that the decay of the reduced bosonic occupation number starting with a pure boson number state shows an initial oscillatory decay and approaches a finite plateau value in the long-time limit. These plateaus are found to increase monotonically with the ratio between the initial boson number and the total number of spins. In the interacting case, we find that the exciton coupling between nearest-neighboring monomers has a significant effect on the photon generation from the excitonic ground state.
\par The rest of the paper is organized as follows. In Sec. II, we will briefly review the diagonalization procedure of the inhomogeneous XX chain with periodic/open boundary conditions, and we introduce the definition of the two types of SOMEs. In Sec. III, we will derive the determinant formulas for the SOMEs using the fermionic technique. Section IV will be devoted to the application of the formalism to the nonlinear optical response of molecular aggregates and to the dynamics of the interacting inhomogeneous Dicke model. Conclusions are drawn in Sec. V.
\section{The XX spin chain and spin-operator matrix elements}
\subsection{The XX spin chain}
\par The XX spin chain in an inhomogeneous transverse magnetic field consists of a chain of $N$ spins 1/2 with nearest-neighbor XX-type interactions, and is given by the Hamiltonian
\begin{eqnarray}\label{XX}
H_{\mathrm{XX}}=\sum^{N}_{j=1}J_j(S^x_jS^x_{j+1}+S^y_jS^y_{j+1})- \sum^N_{j=1}h_j\left(S^z_j+\frac{1}{2}\right),~~~
\end{eqnarray}
where $S^{\alpha}_j$ ($\alpha=x,y, z$) are the spin-1/2 operators, $J_j$ is the (inhomogeneous) isotropic nearest-neighbor coupling between spin-$j$ and spin-$(j+1)$, and $h_j$ is the magnetic field imposed on spin-$j$. We assume periodic boundary condition $S^\alpha_{N+1}=S^\alpha_1$ when $J_N\neq0$. The usual homogeneous periodic (open) XX chain described by $H_{\mathrm{PBC}}(h,J)$ [$H_{\mathrm{OBC}}(h,J)$] can be obtained by setting $J_j=J,~\forall j$ [$J_j=J$, ($j\neq N$); $J_N=0$], and  $h_j=h,~\forall j$. In the noninteracting limit $J_j=0,~\forall j$, $H_{\rm{XX}}$ reduces to the atomic model $H_{\rm{atom}}(\{h_j\})$ used by Dicke in the discussion of the superradiance phenomenon~\cite{Dicke}.
\par To introduce the notations used later for the illustration of our problem, we first briefly review the diagonalization procedures of the Hamiltonian (\ref{XX}). The first step is to perform the Jordan-Wigner transformation (with $S^\pm_j=S^x_j\pm iS^y_j$)
\begin{eqnarray}\label{JWT}
S^+_j=c^\dag_j T_j,~S^-_j=c_j T_j,~S^z_j=c^\dag_j c_j-\frac{1}{2},
\end{eqnarray}
where $c^\dag_j$ creates a spinless fermion at site $j$ and $T_j=\prod^{j-1}_{l=1}(1-2c^\dag_lc_l)$ are the Jordan-Wigner strings, which include $T_{N+1}=e^{i\pi\sum^N_{l=1}c^\dag_lc_l}$ as the fermion parity operator. It can be easily checked that the following relations hold
\begin{equation}
T_{j} c_l T_{j} =
\begin{cases}
c_l &j\leq l,\\
  -c_l& j >l,\\
\end{cases}
	\label{TCT}
\end{equation}
After the Jordan-Wigner transformation, $H_{\mathrm{XX}}$ is mapped into a noninteracting spinless fermion model,
\begin{eqnarray}\label{XX_JWT}
H_{\mathrm{XX}}&=&\frac{1}{2}\sum^{N-1}_{j=1}J_j(c^\dag_jc_{j+1}+c^\dag_{j+1}c_j)-\sum^N_{j=1}h_j c^\dag_jc_j\nonumber\\
&&-\frac{1}{2} J_N(c^\dag_Nc_{1}+c^\dag_{ 1}c_N)T_{N+1},
\end{eqnarray}
where we have separated out the bulk and boundary parts of the hopping terms.
\par For $J_N=0$, we have an open XX chain that can be
brought into a diagonal form
\begin{eqnarray}\label{Dia_open}
H^{(\mathrm{o})}_{\mathrm{XX}}=\sum^{N}_{\eta=1} E_\eta \xi^\dag_\eta \xi_\eta,
\end{eqnarray}
by further performing a canonical transformation of the fermions,
\begin{eqnarray}\label{CanT}
c_j=\sum^{N}_{\eta=1}U_{\eta j}\xi_\eta,~\xi_\eta=\sum^{N}_{j=1}U^*_{\eta j}c_j,
\end{eqnarray}
where $U$ is an $N\times N$ unitary matrix satisfying $\sum_j U_{\eta j}U^*_{\eta'j}=\delta_{\eta\eta'}$. In the special case of a homogeneous open XX chain described by $H_{\rm{OBC}}(h,J)$, the canonical transformation is of the form
\begin{eqnarray}\label{CanT_open}
U_{\eta j}=\sqrt{\frac{2}{N+1}}\sin K_\eta j~\mathrm{with}~K_{\eta}=\frac{\eta\pi}{N+1}.
\end{eqnarray}
The corresponding single particle dispersion is $E_{\eta}=J\cos K_{\eta}-h$.
\par For $J_N\neq0$, since the fermion parity $T_{N+1}$ is conserved, one can separately diagonalize $H_{\mathrm{XX}}$ in the two subspaces with even ($T_{N+1}=1$) and odd ($T_{N+1}=-1$) number of fermions. This can be achieved by two individual sets of canonical transformations
\begin{eqnarray}\label{Can_P}
c_j=\sum^{N}_{\eta=1}U^{(\sigma)}_{\eta j}\xi_{\eta,\sigma},~\xi_{\eta,\sigma}=\sum^{N}_{j=1}U^{(\sigma)*}_{\eta j}c_j,
\end{eqnarray}
which gives the diagonal form
\begin{eqnarray}\label{H-diag}
H^{(\mathrm{p})}_{\mathrm{XX}}&=&\sum_{\sigma=\pm}\frac{1+\sigma T_{N+1}}{2}H^{(\mathrm{p})}_\sigma\frac{1+\sigma T_{N+1}}{2},\nonumber\\
H^{(\mathrm{p})}_\sigma&=&\sum^{N}_{\eta=1}E_{\eta,\sigma} \xi^\dag_{\eta,\sigma}\xi_{\eta,\sigma},
\end{eqnarray}
where $\sigma=1(-1)$ indicates the even (odd) subspace. In the special case of a homogeneous periodic XX chain described by $H_{\rm{PBC}}(h,J)$, the canonical transformation is of the form ($N=$ even)
\begin{eqnarray}\label{FT}
U^{(\sigma)}_{\eta j}=\frac{1}{\sqrt{N}}e^{iK^{(\sigma)}_{\eta}j},
\end{eqnarray}
where
\begin{eqnarray}\label{Keo}
K^{(\sigma)}_{\eta}=-\pi+\left(2\eta+\frac{\sigma-3}{2}\right)\frac{\pi }{N},~\eta=1,2,\cdots,N
\end{eqnarray}
are the allowed wave numbers that give the periodic ($\sigma=-1$) or antiperiodic ($\sigma=+1$) boundary conditions in the $c$-fermion representation.
\par For notational convenience, we define vectors made up of spatial and mode indices $\vec{j}_m\equiv(j_1,j_2,\cdots,j_m)$ and $\vec{\eta}_{m}\equiv(\eta_1,\eta_2,\cdots,\eta_m)$, with the convention $1\leq j_1<j_2<\cdots<j_m\leq N$ and $1\leq\eta_1<\eta_2<\cdots<\eta_m\leq N$. Thus, any eigenstate of $H^{(\mathrm{p})}_{\mathrm{XX}}$ can be written as $
|\vec{\eta}^{(\mathrm{p})}_m\rangle=\prod^m_{l=1}\xi^\dag_{\eta_l,\sigma_m}|0\rangle$, where $|0\rangle$ is the vacuum state of the $c$-fermions and $\sigma_m=+$ ($-$) if $m$ is even (odd). The eigenstate $|\vec{\eta}^{(\mathrm{p})}_m\rangle$ is related to the real-space state $|\vec{j}_m\rangle=\prod^m_{l=1}c^\dag_{j_l}|0\rangle$ filled by $m$ fermions on sites $j_1,j_2,\cdots,j_m$ via
\begin{eqnarray}\label{jnkn}
|\vec{\eta}^{(\mathrm{p})}_m\rangle=\sum_{\vec{j}_m}\mathcal{S}_{\vec{\eta}_m;\vec{j}_m}|\vec{j}_m\rangle,
\end{eqnarray}
where the function
\begin{eqnarray}\label{Slater}
\mathcal{S}_{\vec{\eta}_m;\vec{j}_m}= \det\left(
                                                  \begin{array}{cccc}
                                                    U^{(\sigma_m)}_{\eta_1,j_1}  & U^{(\sigma_m)}_{\eta_1,j_2} & \cdot & U^{(\sigma_m)}_{\eta_1,j_m} \\
                                                    U^{(\sigma_m)}_{\eta_2,j_1}  & U^{(\sigma_m)}_{\eta_2,j_2} & \cdot & U^{(\sigma_m)}_{\eta_2,j_m} \\
                                                    \cdot & \cdot & \cdot & \cdot \\
                                                    U^{(\sigma_m)}_{\eta_m,j_1}  & U^{(\sigma_m)}_{\eta_m,j_2} & \cdot & U^{(\sigma_m)}_{\eta_m,j_m} \\
                                                  \end{array}
                                                \right)
\end{eqnarray}
is the Slater determinant made up of the transformation coefficients. A similar expression to Eq.~(\ref{jnkn}) holds for the eigenstate $|\vec{\eta}^{(\mathrm{o})}_m\rangle=\prod^m_{l=1}\xi^\dag_{\eta_l}|0\rangle$ of $H^{(\mathrm{o})}_{\mathrm{XX}}$. It is useful to observe that the fermion occupation state $|\vec{j}_m\rangle$ is consistent with the real-space Ising configuration $\prod^m_{l=1}S^+_{j_l}|\downarrow\cdots\downarrow\rangle$ in the spin representation:
\begin{eqnarray}\label{Isingconfig}
|\vec{j}_m\rangle=\prod^m_{l=1}S^+_{j_l}|\downarrow\cdots\downarrow\rangle.
\end{eqnarray}
\par In the atomic limit described by $H_{\rm{atom}}(\{h_j\})$, we simply have $|\vec{\eta}_m\rangle=|\vec{j}_m\rangle$. For a uniform magnetic field $h$, all the $C^m_N$ states $\{|\vec{\eta}_m\rangle\}$ are degenerate and process a common energy $\varepsilon_m=-hm$. The fully symmetric Dicke ladder states~\cite{Dicke} are then given by the symmetric linear superpositions of these states,
\begin{eqnarray}\label{Dicke}
|\frac{N}{2},m-\frac{N}{2}\rangle=\frac{1}{\sqrt{C^m_N}}\sum_{\vec{\eta}_m}|\vec{\eta}_m\rangle,
\end{eqnarray}
where $C^m_N=\frac{N!}{m!(N-m)!}$ is the binomial coefficient.
\subsection{The spin-operator matrix element: some known results}
\par In this work, we are interested in the following SOME:
\begin{eqnarray}\label{SOM_def}
F^{\mathrm{( p/o)}}_{j;\vec{\eta}_{n+1},\vec{\chi}_n}\equiv\langle \vec{\chi}^{\mathrm{( p/o)}}_n|S^-_j|\vec{\eta}^{\mathrm{( p/o)}}_{n+1}\rangle,
\end{eqnarray}
between two eigenstates $|\vec{\eta}^{\mathrm{( p)}}_{n+1}\rangle$ and $|\vec{\chi}^{\mathrm{( p)}}_{n}\rangle$ ($|\vec{\eta}^{\mathrm{( o)}}_{n+1}\rangle$ and $|\vec{\chi}^{\mathrm{( o)}}_{n}\rangle$) of $H^{(\mathrm{p})}_{\mathrm{XX}}$ ($H^{(\mathrm{o})}_{\mathrm{XX}}$) that differ by a single excitation, where $S^-_j$ is the lowering operator of spin-$j$. 
We also define the \emph{collective} SOME
\begin{eqnarray}\label{SOM_gj_def}
F^{\mathrm{( p/o)}}_{\vec{\eta}_{n+1},\vec{\chi}_n}(\{g_j\})&\equiv&\langle \vec{\chi}^{\mathrm{( p/o)}}_n|\sum^N_{j=1}g_jS^-_j|\vec{\eta}^{\mathrm{( p/o)}}_{n+1}\rangle\nonumber\\
&=&\sum^N_{j=1}g_jF^{\mathrm{( p/o)}}_{j;\vec{\eta}_{n+1},\vec{\chi}_n},
\end{eqnarray}
which is associated with a distribution of some variables, $\{g_j\}$ ($j=1,2,\cdots,N$), e.g., a nonuniform system-bath coupling configuration~\cite{PRB2016}.
\par The other type of SOMEs we will consider involve two spin operators and two eigenstates $|\vec{\chi}^{\mathrm{( p/o)}}_{n}\rangle$ and $|\vec{\chi}'^{\mathrm{( p/o)}}_{n}\rangle$ with the same number of excitations,
\begin{eqnarray}\label{SOM1_def}
\bar{F}^{\mathrm{( p/o)}}_{l,l';\vec{\chi}_{n},\vec{\chi}'_{n}}=\bar{F}^{\mathrm{( p/o)}*}_{l',l;\vec{\chi}'_{n},\vec{\chi}_{n}}\equiv\langle \vec{\chi}^{\mathrm{( p/o)}}_{n}|S^-_{l'}S^+_l|\vec{\chi}'^{\mathrm{( p/o)}}_{n}\rangle,
\end{eqnarray}
which is relevant to, for example, the superradiance master equation describing linear molecular aggregates interacting with a light field~\cite{Hanamura}. In the atomic limit, the energy basis $\{|\vec{\chi}^{\mathrm{( p/o)}}_n\rangle\}$ reduces to the Ising configurations $\{|\vec{j}_n\rangle\}$, so that $\bar{F}^{\mathrm{( p/o)}}_{l,l';\vec{j}_{n},\vec{j}'_{n}}$ gives the matrix element of the $XY$-type spin interaction $S^x_lS^x_{l'}+S^y_lS^y_{l'}$ in real-space (for $l\neq l'$),
\begin{eqnarray}\label{SXY}
&&\langle \vec{j}_n|S^x_lS^x_{l'}+S^y_lS^y_{l'}|\vec{j}'_n\rangle=\frac{1}{2}\left(\bar{F}^{\mathrm{( p/o)}}_{l,l';\vec{j}_{n},\vec{j}'_{n}}+\bar{F}^{\mathrm{( p/o)}}_{l',l;\vec{j}_{n},\vec{j}'_{n}}\right).\nonumber\\
\end{eqnarray}

\par At first glance, it seems difficult to calculate $F^{\mathrm{( p/o)}}_{j;\vec{\eta}_{n+1},\vec{\chi}_n}$ or $\bar{F}^{\mathrm{( p/o)}}_{l,l';\vec{\chi}_{n},\vec{\chi}'_{n}}$ due to the Jordan-Wigner string involved in the spin operators. A naive attempt is to write the eigenstates $|\vec{\eta}^{\mathrm{( p/o)}}_{n+1}\rangle$ in terms of the real-space fermion states through Eq.~(\ref{jnkn}) and rearrange the operators using fermion commutation rules. As shown in Refs.~\cite{Spano1991,Hanamura} for a uniform distribution $g_j=g$ (and independently in Refs.~\cite{PRA2014,PRB2016} for a nonuniform distribution $\{g_j\}$), the matrix element $F^{\mathrm{( p)}}_{\vec{\eta}_{n+1},\vec{\chi}_n}(\{g_j\})$ for the periodic (not necessarily homogeneous) XX chain can indeed be expressed in terms of the Slater determinants as
 \begin{eqnarray}\label{f_SS}
F^{\mathrm{( p)}}_{\vec{\eta}_{n+1},\vec{\chi}_n}(\{g_j\})&=& \sum_{\vec{j}_{n+1}}\mathcal{S}_{\vec{\eta}_{n+1};\vec{j}_{n+1}} \sum^{n+1}_{l=1}g_{j_l}\mathcal{S}^*_{\vec{\chi}_n;\vec{j}^{(l)}_{n+1}},
\end{eqnarray}
where the vector $\vec{j}^{(l)}_{n+1}=(j_1,...,j_{l-1},j_{l+1},...,j_{n+1})$ is a string of length $n$ with the element $j_l$ being removed from the $(n+1)$-string $\vec{j}_{n+1}$. For the sake of completeness, in Appendix \ref{AppA} we give some details of the derivation of Eq.~(\ref{f_SS}) under periodic boundary conditions (similar expressions hold for the open boundary conditions).
\par However, Eq.~(\ref{f_SS}) still looks cumbersome to calculate~\cite{Spano1991,PRB2016} due to the multisums over the $n+1$ site indices $j_1<j_2<\cdots<j_{n+1}$. For the simpler case with a uniform distribution $g=g_j$, as well as a periodic homogeneous XX chain described by $H_{\rm{PBC}}$, the matrix elements $F^{\mathrm{(\rm{PBC})}}_{\vec{\eta}_{n+1},\vec{\chi}_n}(g)$ do admit closed forms~\cite{Hanamura},
\begin{eqnarray}\label{DI}
F^{\mathrm{(\rm{PBC})}}_{\vec{\eta}_{n+1},\vec{\chi}_n} (g)&=&g2^nN^{\frac{1}{2}-n} \delta\left(\sum^{n+1}_{j=1}K^{(\sigma_{n+1})}_{\eta_j},\sum^n_{i=1}K^{(\sigma_n)}_{\chi_i}\right)\cdot\nonumber\\
&&h_{\vec{\eta}_{n+1}; \vec{\chi}_n},
\end{eqnarray}
where the Kronecker delta-function $\delta(x,y)$ is $1$ when $x=y+2\pi m$ ($m\in Z$), and $0$ otherwise, and $h_{\vec{\eta}_{n+1}; \vec{\chi}_n}$ is a factorized function of the momentum configurations:
\begin{eqnarray}\label{DI_h}
&&h_{\vec{\eta}_{n+1}; \vec{\chi}_n}\nonumber\\
&=&\frac{\prod_{i>i'}(e^{-iK^{(\sigma_n)}_{\chi_i}}-e^{-iK^{(\sigma_n)}_{\chi_{i'}}})\prod_{j>j'}(e^{iK^{(\sigma_{n+1})}_{\eta_j}}-e^{iK^{(\sigma_{n+1})}_{\eta_{j'}}})}{\prod^n_{i=1}\prod^{n+1}_{j=1}(1-e^{-i(K^{(\sigma_{n+1})}_{\eta_j}-K^{(\sigma_{n})}_{\chi_i})})}.\nonumber\\
\end{eqnarray}
As claimed by the authors of Ref.~\cite{Hanamura}, Eq.~(\ref{DI}) can be obtained after ``some tedious calculation"~\cite{fnote}. In spite of some attempts to derive Eq.~(\ref{DI}) directly by using the properties of determinants (see, e.g., Ref.~\cite{math}), efficient evaluation of $F^{\mathrm{( p)}}_{\vec{\eta}_{n+1},\vec{\chi}_n}(\{g_j\})$ beyond Eq.~(\ref{f_SS}) is still absent.

\section{Evaluation of the spin-operator matrix elements: fermionic approach}\label{proof}

\par Instead of calculating $F^{\mathrm{( p)}}_{\vec{\eta}_{n+1},\vec{\chi}_n}(\{g_j\})$ directly from Eq.~(\ref{f_SS}), in this section we follow a different strategy by starting with the definition of the SOMEs, Eqs.~(\ref{SOM_def}) and (\ref{SOM1_def}), and we derive simple expressions of them by employing a fermionic approach developed in the work of Iorgov \emph{et al.}~\cite{Iorgov} in the derivation of factorized expressions for the SOMEs in the quantum Ising chain. We will show that either $F^{\mathrm{( p/o)}}_{j;\vec{\eta}_{n+1},\vec{\chi}_n}$ or $\bar{F}^{\mathrm{( p/o)}}_{l,l';\vec{\chi}_{n},\vec{\chi}'_{n}}$ can be expressed as the determinant of some $(n+1)\times(n+1)$ square matrix involving the coefficients of the corresponding canonical transformations. In turn, Eq.~(\ref{DI}) is shown to be a direct consequence of the application of the obtained general formulas to a homogeneous periodic XX chain with uniform system-bath coupling. In the following, we will focus on the periodic XX chain since similar results hold for the open XX chain.
\subsection{Calculation of $F^{\mathrm{( p)}}_{j;\vec{\eta}_{n+1},\vec{\chi}_n}$}\label{sectionIIA}
\par We first start with the complex conjugate of Eq.~(\ref{SOM_def}),
\begin{eqnarray}\label{SOM_cc}
&&F^{\mathrm{( p)*}}_{j;\vec{\eta}_{n+1},\vec{\chi}_n}\equiv\langle\vec{\eta}^{\mathrm{( p)}}_{n+1}|S^+_j|\vec{\chi}^{\mathrm{( p)}}_n\rangle\nonumber\\
&=&\langle0|\xi_{\eta_{n+1},\bar{\sigma}_{n}}\cdots \xi_{\eta_{1},\bar{\sigma}_{n}}T_jc^\dag_j \xi^\dag_{\chi_1,\sigma_n}\cdots \xi^\dag_{\chi_n,\sigma_n}|0\rangle\nonumber\\
&=&\sum^N_{\chi=1}U^{(\sigma_n)*}_{\chi,j}\langle0|\xi_{\eta_{n+1},\bar{\sigma}_{n}}\cdots \xi_{\eta_{1},\bar{\sigma}_{n}}T_j\xi^\dag_{\chi,\sigma_n} \xi^\dag_{\chi_1,\sigma_n}\cdots \xi^\dag_{\chi_n,\sigma_n}|0\rangle,\nonumber\\
\end{eqnarray}
where $\bar{\sigma}=-\sigma$ and we have used $S^+_j=T_j c^\dag_j$ and shifted to the energy representation. We emphasize that the mode index $\chi$ is not necessarily less than $\chi_1$ in Eq.~(\ref{SOM_cc}). Let us focus on the expectation value in the last line of Eq.~(\ref{SOM_cc}),
\begin{eqnarray}\label{Dj}
&&D^{(j)}_{\eta_1,\cdots,\eta_{n+1};\chi,\chi_1,\cdots,\chi_n}\nonumber\\
&\equiv&\langle0|\xi_{\eta_{n+1},\bar{\sigma}_{n}}\cdots \xi_{\eta_{1},\bar{\sigma}_{n}}T_j\xi^\dag_{\chi,\sigma_n} \xi^\dag_{\chi_1,\sigma_n}\cdots \xi^\dag_{\chi_n,\sigma_n}|0\rangle.
\end{eqnarray}
The trick is to insert the identity $T_jT_j=1$ between $\xi_{\eta_{2},\bar{\sigma}_{n}}$ and $\xi_{\eta_{1},\bar{\sigma}_{n}}$ in Eq.~(\ref{Dj}):
\begin{eqnarray}\label{Dj1}
&&D^{(j)}_{\eta_1,\cdots,\eta_{n+1};\chi,\chi_1,\cdots,\chi_n}\nonumber\\
&=& \langle0|\xi_{\eta_{n+1},\bar{\sigma}_{n}}\cdots \xi_{\eta_{2},\bar{\sigma}_{n}}T_j(T_j\xi_{\eta_{1},\bar{\sigma}_{n}}T_j)\xi^\dag_{\chi,\sigma_n} \xi^\dag_{\chi_1,\sigma_n}\cdots \xi^\dag_{\chi_n,\sigma_n}|0\rangle\nonumber\\
&=&\sum^N_{n'=1}U^{(\sigma_{n+1})*}_{\eta_1,n'} \langle0|\xi_{\eta_{n+1},\bar{\sigma}_{n}}\cdots \xi_{\eta_{2},\bar{\sigma}_{n}}T_j(T_jc_{n'}T_j)\nonumber\\
&&\xi^\dag_{\chi,\sigma_n} \xi^\dag_{\chi_1,\sigma_n}\cdots \xi^\dag_{\chi_n,\sigma_n}|0\rangle\nonumber\\
&=&\sum^N_{\chi'=1}A^{(j),(\sigma_n)}_{\eta_1,\chi'}\langle0|\xi_{\eta_{n+1},\bar{\sigma}_{n}}\cdots \xi_{\eta_{2},\bar{\sigma}_{n}}T_j\xi_{\chi',\sigma_n} \nonumber\\
&&\xi^\dag_{\chi,\sigma_n} \xi^\dag_{\chi_1,\sigma_n}\cdots \xi^\dag_{\chi_n,\sigma_n}|0\rangle,
\end{eqnarray}
where the coefficients $A^{(j),(n)}_{\eta,\sigma}$ are given by
\begin{eqnarray}\label{Apk}
A^{(j),(\sigma)}_{\eta,\chi}&=&A^{(j),(\bar{\sigma})*}_{\chi,\eta}\equiv\left(\sum^N_{l=1}-2\sum^{j-1}_{l=1}\right) U^{(\bar{\sigma})*}_{\eta,l}U^{(\sigma)}_{\chi,l},
\end{eqnarray}
and we have used Eq.~(\ref{TCT}) in the derivation of the last line of Eq.~(\ref{Dj1}). It can be easily checked that
\begin{equation}
\sum^N_{\chi=1}A^{(j),(\sigma)*}_{\eta ,\chi}U^{(\sigma)}_{\chi,j'}=
\begin{cases}
 \mathrm{sgn}(j'-j)U^{(\bar{\sigma})}_{\eta,j'} &j'\neq j,\\
U^{(\bar{\sigma})}_{\eta,j}& j=j'.\\
\end{cases}
	\label{AU_U}
\end{equation}
The $A^{(j),(\sigma)}_{\eta ,\chi}$'s can be combined to form
\begin{eqnarray}\label{sum_AA}
\bar{A}^{(j,j'),(\sigma)}_{\eta,\eta'}&\equiv&\sum^N_{\chi=1}A^{(j),(\sigma)}_{\eta,\chi}A^{(j'),(\sigma)*}_{\eta',\chi}=\sum^N_{\chi=1}A^{(j),(\bar{\sigma})*}_{\chi,\eta}A^{(j'),(\bar{\sigma})}_{\chi,\eta'}\nonumber\\
&=&\delta_{\eta,\eta'}-2 \sum^{j_{\max}-1}_{l=j_{\min}} U^{(\bar{\sigma})*}_{\eta,l} U^{(\bar{\sigma})}_{\eta',l},
\end{eqnarray}
where $j_{\min}=\min\{j,j'\}$ and $j_{\max}=\max\{j,j'\}$, respectively. Equation (\ref{sum_AA}) will be used below to derive the expression for $\bar{F}^{\mathrm{( p)}}_{l,l';\vec{\chi}_{n},\vec{\chi}'_{n}}$. Note that
\begin{eqnarray}\label{A_A_A}
\bar{A}^{(j,j'),(\sigma)}_{\eta,\eta'}=\bar{A}^{(j',j),(\sigma)}_{\eta,\eta'}=\bar{A}^{(j,j'),(\sigma)*}_{\eta',\eta}.
\end{eqnarray}
\par Let us now come back to Eq.~(\ref{Dj1}) and note that only $\chi'=\chi,\chi_1,\cdots,\chi_n$ contribute in its last line, we hence have
\begin{eqnarray}\label{D_LP}
&&D^{(j)}_{\eta_1,\cdots,\eta_{n+1};\chi,\chi_1,\cdots,\chi_n}= A^{(j),(\sigma_n)}_{\eta_1,\chi} D^{(j)}_{\eta_2,\cdots,\eta_{n+1};\chi_1,\cdots,\chi_n}\nonumber\\
&&+ \sum^n_{m=1}(-1)^mA^{(j),(\sigma_n)}_{\eta_1,\chi_m} D^{(j)}_{\eta_2,\cdots,\eta_{n+1};\chi,\chi_1,\cdots,\chi_{m-1},\chi_{m+1},\cdots,\chi_n},\nonumber\\
\end{eqnarray}
where the factor $(-1)^m$ arises from moving $\xi_{\chi',\sigma_n}$ to the right to pass by the $m$ creation operators $\xi^\dag_{\chi,\sigma_n}$, $\xi^\dag_{\chi_1,\sigma_n},\cdots$, and $\xi^\dag_{\chi_{m-1},\sigma_n}$. The expansion on the right-hand side of Eq.~(\ref{D_LP}) reminds us of the Laplace expansion of a determinant. By noting that $\langle0|T_j|0\rangle=1$, we thus obtain
\begin{eqnarray}\label{CME3}
&&D^{(j)}_{\eta_1,\cdots,\eta_{n+1};\chi,\chi_1,\cdots,\chi_n}\nonumber\\
&=& \det\left(
       \begin{array}{ccccc}
         A^{(j),(\sigma_n)}_{\eta_1,\chi} & A^{(j),(\sigma_n)}_{\eta_1,\chi_1} & A^{(j),(\sigma_n)}_{\eta_1,\chi_2} & \cdot & A^{(j),(\sigma_n)}_{\eta_1,\chi_n} \\
          A^{(j),(\sigma_n)}_{\eta_2,\chi} & A^{(j),(\sigma_n)}_{\eta_2,\chi_1} & A^{(j),(\sigma_n)}_{\eta_2,\chi_2} & \cdot & A^{(j),(\sigma_n)}_{\eta_2,\chi_n} \\
         \cdot & \cdot & \cdot & \cdot & \cdot \\
         \cdot & \cdot & \cdot & \cdot & \cdot \\
          A^{(j),(\sigma_n)}_{\eta_{n+1},\chi} & A^{(j),(\sigma_n)}_{\eta_{n+1},\chi_1} & A^{(j),(\sigma_n)}_{\eta_{n+1},\chi_2} & \cdot & A^{(j),(\sigma_n)}_{\eta_{n+1},\chi_n} \\
       \end{array}
     \right).~~~~~~~
\end{eqnarray}
Substituting this equation into Eq.~(\ref{SOM_cc}), and using Eq.~(\ref{AU_U}), we finally have
\begin{eqnarray}\label{SOM_cc_final}
&&F^{\mathrm{( p)}}_{j;\vec{\eta}_{n+1},\vec{\chi}_n}\nonumber\\
&=&\det\left(
       \begin{array}{ccccc}
        U^{(\bar{\sigma}_n)}_{\eta_1,j} & A^{(j),(\sigma_n)*}_{\eta_1,\chi_1} & A^{(j),(\sigma_n)*}_{\eta_1,\chi_2} & \cdot & A^{(j),(\sigma_n)*}_{\eta_1,\chi_n} \\
         U^{(\bar{\sigma}_n)}_{\eta_2,j} & A^{(j),(\sigma_n)*}_{\eta_2,\chi_1} & A^{(j),(\sigma_n)*}_{\eta_2,\chi_2} & \cdot & A^{(j),(\sigma_n)*}_{\eta_2,\chi_n} \\
         \cdot & \cdot & \cdot & \cdot & \cdot \\
         \cdot & \cdot & \cdot & \cdot & \cdot \\
        U^{(\bar{\sigma}_n)}_{\eta_{n+1},j} & A^{(j),(\sigma_n)*}_{\eta_{n+1},\chi_1} & A^{(j),(\sigma_n)*}_{\eta_{n+1},\chi_2} & \cdot & A^{(j),(\sigma_n)*}_{\eta_{n+1},\chi_n} \\
       \end{array}
     \right).\nonumber\\
\end{eqnarray}
Equation~(\ref{SOM_cc_final}) is one of the main results of this paper, and it provides an easy way to numerically calculate the SOMEs of $S^-_j$ for an inhomogeneous XX chain by evaluating a single determinant. Since there are no general formulas to evaluate the sum of determinants, calculating the $C^n_{N}C^{n+1}_N$ collective SOMEs $F_{\vec{\eta}_{n+1},\vec{\chi}_n}$ given by Eq.~(\ref{SOM_gj_def}) requires the calculation of all the $NC^n_{N}C^{n+1}_N$ SOMEs $F_{j;\vec{\eta}_{n+1},\vec{\chi}_n}$, which costs $N$ times the memory needed for the former. Note that Eqs.~(\ref{Dj})-(\ref{SOM_cc_final}) also hold for the open XX chain, with the indices $\sigma$'s removed from the corresponding expressions. In the atomic limit, it is easy to show that Eq.~(\ref{SOM_cc_final}) reduces to $F^{\mathrm{( p)}}_{j;\vec{l}_{n+1},\vec{l}'_n}=\sum^{n+1}_{m=1} \delta_{l_m,j}\delta_{\vec{l}^{(m)}_{n+1},\vec{l}'_n}$ for two Ising configurations $|\vec{l}_{n+1}\rangle$ and $|\vec{l}'_n\rangle$, which is consistent with the definition $F^{\mathrm{( p)}}_{j;\vec{l}_{n+1},\vec{l}'_n}=\langle \vec{l}'_n|S^-_j|\vec{l}_{n+1}\rangle$.
\subsection{The homogeneous XX chains}
\par For a homogeneous open XX chain described by $H_{\rm{OBC}}$, we have (with $ \sin\alpha x/\sin x=\alpha$ when $x=0$)
\begin{eqnarray}\label{A_tilde}
&&A^{(j)}_{\eta,\chi}= \delta_{\eta,\chi}-\frac{1}{N+1}\nonumber\\
&&\left[\frac{ \sin(j-\frac{1}{2})(K_{\eta}-K_{\chi})}{\sin \frac{1}{2}(K_{\eta}-K_{\chi})}-\frac{ \sin(j-\frac{1}{2})(K_{\eta}+K_{\chi})}{\sin \frac{1}{2}(K_{\eta}+K_{\chi})}\right],\nonumber\\
\end{eqnarray}
and hence
\begin{eqnarray}\label{SOM_cc_open_1}
&&F^{\rm{(OBC)}}_{j;\vec{\eta}_{n+1},\vec{\chi}_{n}}=\sqrt{\frac{2}{N+1}}\nonumber\\
&&\det\left(
       \begin{array}{ccccc}
         \sin K_{\eta_1}j & A^{(j)}_{ \eta_1 , \chi_1 } & A^{(j)}_{\eta_1,\chi_2} & \cdot & A^{(j)}_{\eta_1,\chi_n} \\
          \sin K_{\eta_2}j & A^{(j)}_{\eta_2,\chi_1} & A^{(j)}_{\eta_2,\chi_2} & \cdot & A^{(j)}_{\eta_2,\chi_n} \\
         \cdot & \cdot & \cdot & \cdot & \cdot \\
         \cdot & \cdot & \cdot & \cdot & \cdot \\
         \sin K_{\eta_{n+1}}j & A^{(j)}_{\eta_{n+1},\chi_1} & A^{(j)}_{\eta_{n+1},\chi_2} & \cdot & A^{(j)}_{\eta_{n+1},\chi_n} \\
       \end{array}
     \right).~~~~~~
\end{eqnarray}
The corresponding collective SOMEs are then calculated by using Eq.~(\ref{SOM_gj_def}).
\par For the homogeneous periodic XX chain considered in Refs.~\cite{Hanamura,PRA2014,PRB2016}, it is easy to check that
\begin{eqnarray}\label{Apk_PBC}
A^{(j),(\sigma_n)}_{\eta,\chi}&=&\frac{2}{N} \frac{ e^{i(K^{(\sigma_{n})}_\chi-K^{(\bar{\sigma}_{n})}_\eta)j} e^{ iK^{(\bar{\sigma}_{n})}_\eta}  }{ e^{iK^{(\bar{\sigma}_{n})}_\eta}-e^{ iK^{(\sigma_{n})}_\chi}}.
\end{eqnarray}
Combining the above equation with Eqs.~(\ref{FT}) and (\ref{SOM_cc_final}), we have
\begin{eqnarray}\label{SOM_cc_final_PBC}
&&F^{\mathrm{(PBC)}*}_{j;\vec{\eta}_{n+1},\vec{\chi}_n}=\frac{1}{\sqrt{N}}\left(\frac{2}{N}\right)^n e^{-i\Delta_{\vec{\eta}_{n+1},\vec{\chi}_n}j}\cdot\det\nonumber\\
    &&\left(
       \begin{array}{ccccc}
         1 & \frac{1}{1-e^{-i(K^{(\bar{\sigma}_{n})}_{\eta_1}-K^{(\sigma_{n})}_{\chi_1})}}&  \cdot \cdot &  \frac{1}{1-e^{-i(K^{(\bar{\sigma}_{n})}_{\eta_1}-K^{(\sigma_{n})}_{\chi_n})}} \\
        1 & \frac{1}{1-e^{-i(K^{(\bar{\sigma}_{n})}_{\eta_2}-K^{(\sigma_{n})}_{\chi_1})}}&  \cdot \cdot &  \frac{1}{1-e^{-i(K^{(\bar{\sigma}_{n})}_{\eta_2}-K^{(\sigma_{n})}_{\chi_n})}} \\
          \cdot \cdot &   \cdot\cdot & \cdot \cdot &  \cdot\cdot \\
  1 & \frac{1}{1-e^{-i(K^{(\bar{\sigma}_{n})}_{\eta_{n+1}}-K^{(\sigma_{n})}_{\chi_1})}}&  \cdot \cdot &  \frac{1}{1-e^{-i(K^{(\bar{\sigma}_{n})}_{\eta_{n+1}}-K^{(\sigma_{n})}_{\chi_n})}} \\
       \end{array}
     \right), ~~~
\end{eqnarray}
where we introduced the momentum transfer between the two states $|\vec{\eta}^{\mathrm{(p)}}_{n+1}\rangle$ and $|\vec{\chi}^{\mathrm{(p)}}_{n}\rangle$,
\begin{eqnarray}\label{dmomentum}
\Delta_{\vec{\eta}_{n+1},\vec{\chi}_n}\equiv\sum^{n+1}_{j=1}K^{(\sigma_{n+1})}_{\eta_j}-\sum^n_{i=1}K^{(\sigma_{n})}_{\chi_i},
\end{eqnarray}
which clearly lies in the set $\{K^{(-)}_\eta\}$.
\par Equation (\ref{SOM_cc_final_PBC}) can be simplified further by noting that the determinant appearing in Eq.~(\ref{SOM_cc_final_PBC}) can be evaluated analytically. In fact, if we set $x_j= e^{-iK^{(\sigma_{n+1})}_{\eta_j}}$ ($j=1,2,\cdots,n+1$), $y_i=e^{iK^{(\sigma_{n})}_{\chi_i}}$ ($i=1,2,\cdots,n$), and $y_0=0$, then the determinant has the form of a variant of the Cauchy determinant,
\begin{eqnarray}\label{cauchy}
\det\frac{1}{1-x_j y_i}=\frac{\prod_{j>j'}(x_j-x_{j'})\prod_{i>i'}(y_i-y_{i'})}{\prod_i\prod_j(1-x_i y_j)},
\end{eqnarray}
which after some manipulation leads to
\begin{eqnarray}\label{CME_final1}
&&F^{\mathrm{(PBC)}*}_{j;\vec{\eta}_{n+1},\vec{\chi}_n} =\frac{1}{\sqrt{N}}\left(\frac{2}{N}\right)^n  e^{i(n-j)\Delta_{\vec{\eta}_{n+1},\vec{\chi}_n}}h^*_{\vec{\eta}_{n+1}; \vec{\chi}_n},\nonumber\\
\end{eqnarray}
where $h_{\vec{\eta}_{n+1}; \vec{\chi}_n}$ is given by Eq.~(\ref{DI_h}).
\par The corresponding collective SOMEs for an inhomogeneous distribution $\{g_j\}$ can be readily calculated by introducing the Fourier transform of $\{g_j\}$
\begin{eqnarray}\label{FTg}
g_j&=&\frac{1}{N}\sum_q e^{-iqj}\tilde{g}_q,\nonumber\\
\tilde{g}_q&=&\tilde{g}^*_{-q}=\sum_j e^{iqj}g_j,~q\in \{K^{(-)}_\eta\},
\end{eqnarray}
which results in
\begin{eqnarray}\label{CME_final2}
&&F^{\mathrm{(PBC)}}_{\vec{\eta}_{n+1},\vec{\chi}_n}(\{g_j\})\nonumber\\
&=& \frac{ \tilde{g}_{\Delta_{\vec{\eta}_{n+1},\vec{\chi}_n}} e^{-i n\Delta_{\vec{\eta}_{n+1},\vec{\chi}_n}}}{\sqrt{N}}\left(\frac{2}{N}\right)^n  h_{\vec{\eta}_{n+1},\vec{\chi}_{n}}.
\end{eqnarray}
The above equation states that for two eigenstates $|\vec{\eta}^{\mathrm{(p)}}_{n+1}\rangle$ and $|\vec{\chi}^{\mathrm{(p)}}_{n}\rangle$ of the homogeneous periodic XX chain with momentum difference $\Delta_{\vec{\eta}_{n+1},\vec{\chi}_n}$, the matrix element of the operator $\sum_j g_jS^-_j$ between the two is simply proportional to the Fourier transform $\tilde{g}_q$ of the distribution $\{g_j\}$ in the mode $q=\Delta_{\vec{\eta}_{n+1},\vec{\chi}_n}$. For a uniform distribution $g_j=g$, we simply have $\tilde{g}_{\Delta_{\vec{\eta}_{n+1},\vec{\chi}_n}}=gN\delta(\Delta_{\vec{\eta}_{n+1},\vec{\chi}_n},0)$, so that Eq.~(\ref{DI}) is recovered.
\par We point out that in Ref.~\cite{PRB2016} the inhomogeneous collective SOMEs $F^{\mathrm{(PBC)}}_{\vec{\eta}_{n+1},\vec{\chi}_n}(\{g_j\})$ were numerically computed by directly using Eq.~(\ref{f_SS}), which is numerically expensive and memory-demanding due to the multisums over the spatial indices $\vec{j}_{n+1}$. The factorized expression Eq.~(\ref{CME_final2}) derived in this work provides an easy way to calculate $F^{\mathrm{(PBC)}}_{\vec{\eta}_{n+1},\vec{\chi}_n}(\{g_j\})$. In addition, the evaluation of $F^{\mathrm{(PBC)}}_{\vec{\eta}_{n+1},\vec{\chi}_n}(\{g_j\})$ does not involve the calculation of the $NC^n_NC^{n+1}_N$ SOMEs $F^{\mathrm{(PBC)}}_{j;\vec{\eta}_{n+1},\vec{\chi}_n}$, thus the computational memory and time can be further saved.
\subsection{Calculation of $\bar{F}^{\mathrm{( p)}}_{l,l';\vec{\chi}_{n},\vec{\chi}'_{n}}$}
\par We now turn to the SOME $\bar{F}^{\mathrm{( p)}}_{l,l';\vec{\chi}_{n},\vec{\chi}'_{n}}$, which can also be evaluated by using a similar fermionic approach. For the sake of simplicity, we outline in Appendix~\ref{AppB} details of the derivation of $\bar{F}^{\mathrm{( p)}}_{l,l';\vec{\chi}_{n},\vec{\chi}'_{n}}$ , which actually closely follow the method used in Sec.~\ref{sectionIIA}.
\par It turns out that $\bar{F}^{\mathrm{( p)}}_{l,l';\vec{\chi}_{n},\vec{\chi}'_{n}}$ is given by
\begin{eqnarray}\label{s_ij_1}
&&\bar{F}^{\mathrm{( p)}}_{l,l';\vec{\chi}_{n},\vec{\chi}'_{n}}=(-1)^{1+\delta_{l,l'}}\cdot\nonumber\\
     &&\det\left(
       \begin{array}{ccccc}
      \delta_{l,l'} &   U^{(\sigma_n)*}_{\chi_1,l} &   U^{(\sigma_n)*}_{\chi_2,l} & \cdot &   U^{(\sigma_n)*}_{\chi_n,l} \\
       U^{(\sigma_n)}_{\chi'_1,l'} &\bar{A}^{(l,l'),(\bar{\sigma}_n)}_{\chi_1,\chi'_1}& \bar{A}^{(l,l'),(\bar{\sigma}_n)}_{\chi_2,\chi'_1} & \cdot &\bar{A}^{(l,l'),(\bar{\sigma}_n)}_{\chi_n,\chi'_1} \\
         \cdot & \cdot & \cdot & \cdot & \cdot \\
         \cdot & \cdot & \cdot & \cdot & \cdot \\
      U^{(\sigma_n)}_{\chi'_n,l'}  &\bar{A}^{(l,l'),(\bar{\sigma}_n)}_{\chi_1,\chi'_n}& \bar{A}^{(l,l'),(\bar{\sigma}_n)}_{\chi_2,\chi'_n} & \cdot &\bar{A}^{(l,l'),(\bar{\sigma}_n)}_{\chi_n,\chi'_n} \\
       \end{array}
 \right).\nonumber\\
\end{eqnarray}
Equation (\ref{s_ij_1}) states that $\bar{F}^{\mathrm{( p)}}_{l,l';\vec{\chi}_{n},\vec{\chi}'_{n}}$ can also be expressed as the determinant of a single $(n+1)\times(n+1)$ matrix. It is worth mentioning that another, more natural way of calculating $\bar{F}^{\mathrm{( p)}}_{l,l';\vec{\chi}_{n},\vec{\chi}'_{n}}$ is to insert the completeness relation $\sum^{N-1}_{m=0}\sum_{\vec{\eta}_{m+1}}|\vec{\eta}^{\mathrm{(p)}}_{m+1}\rangle\langle \vec{\eta}^{\mathrm{(p)}}_{m+1}|=1$ between $S^-_{l'}$ and $S^+_l$ in the definition given by Eq.~(\ref{SOM1_def}), and we observe that only those states with $n+1$ excitations contribute~\cite{Hanamura1995}. It is shown in Appendix~\ref{AppC} that a combination of the obtained results for $F^{\mathrm{( p)}}_{j;\vec{\eta}_{n+1},\vec{\chi}_n}$ [Eq.~(\ref{SOM_cc_final})] with the Cauchy-Binet formula can also yield Eq.~(\ref{s_ij_1}).
\par Let us now look into the case of $l=l'$ a little bit further. In this case,
\begin{eqnarray}\label{s_ijll}
&&\bar{F}^{\mathrm{( p)}}_{l,l;\vec{\chi}_{n},\vec{\chi}'_{n}}\nonumber\\
     &=&\det\left(
       \begin{array}{ccccc}
     1 &   U^{(\sigma_n)*}_{\chi_1,l} &   U^{(\sigma_n)*}_{\chi_2,l} & \cdot &   U^{(\sigma_n)*}_{\chi_n,l} \\
       U^{(\sigma_n)}_{\chi'_1,l } &\delta_{\chi_1,\chi'_1}& \delta_{\chi_2,\chi'_1} & \cdot &\delta_{\chi_n,\chi'_1} \\
         \cdot & \cdot & \cdot & \cdot & \cdot \\
         \cdot & \cdot & \cdot & \cdot & \cdot \\
      U^{(\sigma_n)}_{\chi'_n,l }  &\delta_{\chi_1,\chi'_n}&\delta_{\chi_2,\chi'_n} & \cdot &\delta_{\chi_n,\chi'_n} \\
       \end{array}
 \right).
\end{eqnarray}
In physically relevant cases with $n\geq 2$, if the set $\{\vec{\chi}_n\}\equiv\{\chi_1,\cdots,\chi_n\}$ contains more than one element, for example $\chi_{\alpha_1},~\chi_{\alpha_2},~\cdots$ not being equal to any element in the set $\{\vec{\chi}'_n\}=\{\chi'_1,\cdots,\chi'_n\}$, then it is obvious that $\bar{F}^{\mathrm{( p)}}_{l,l;\vec{\chi}_{n},\vec{\chi}'_{n}}$ vanishes since the $(\alpha_1+1)$th, $(\alpha_2+1)$th, $\cdots$ columns of the matrix in the above equation have their last $n$ elements being zero. Thus, the necessary condition for $\bar{F}^{\mathrm{( p)}}_{l,l;\vec{\chi}_{n},\vec{\chi}'_{n}}$ being finite is that the two sets $\{\vec{\chi}_n\}$ and $\{\vec{\chi}'_n\}$ share either $n$ or $n-1$ elements. Actually, by performing a Laplace expansion of the determinant in Eq.~(\ref{s_ijll}) along the first row, we arrive at
\begin{eqnarray}\label{s_ijllLP}
&&\bar{F}^{\mathrm{( p)}}_{l,l;\vec{\chi}_{n},\vec{\chi}'_{n}}=\delta_{\vec{\chi}_n,\vec{\chi}'_{n}} \nonumber\\
     && -\sum^n_{\alpha=1} \sum^n_{\alpha'=1}(-1)^{\alpha+\alpha'}  U^{(\sigma_{n})*}_{\chi_\alpha,l}  U^{(\sigma_{n}) }_{\chi'_{\alpha'},l}   \delta_{\vec{\chi}^{(\alpha)}_n,\vec{\chi}'^{(\alpha')}_n}.
\end{eqnarray}
In the double summation on the right-hand side of the above equation, it is easy to see that $\chi_\alpha\neq \chi'_{\gamma'}$ when $\gamma'\neq\alpha'$, and $\chi'_{\alpha'}\neq \chi_{\gamma}$ when $\gamma\neq\alpha$. Thus, $\chi_\alpha=\chi'_{\alpha'}$ corresponds to the case of $\vec{\chi}_n=\vec{\chi}'_n$, and $\chi_\alpha\neq\chi'_{\alpha'}$ corresponds to the case that $\{\vec{\chi}_n\}$ and $\{\vec{\chi}'_n\}$ share exactly $n-1$ elements, i.e.,
\begin{eqnarray}
&&(\chi_1,\cdots,\chi_{\alpha-1},\chi_{\alpha+1},\cdots,\chi_n)\nonumber\\
&=&(\chi'_1,\cdots,\chi'_{\alpha'-1},\chi'_{\alpha'+1},\cdots,\chi'_n),
\end{eqnarray}
with $\chi_{\alpha}\notin \{\vec{\chi}'_n\}$ and $\chi'_{\alpha'}\notin \{\vec{\chi}_n\}$, respectively.
\par For $\vec{\chi}_n=\vec{\chi}'_n$, we have
\begin{eqnarray}\label{s_ij_1ll2}
\bar{F}^{\mathrm{( p)}}_{l,l;\vec{\chi}_{n},\vec{\chi}_{n}}&=&1-\sum^n_{\alpha=1}U^{(\sigma_{n})}_{\chi_\alpha,l}U^{(\sigma_{n})*}_{\chi_\alpha,l}.
\end{eqnarray}
In the latter case, one can show that, among the totally $C^n_{N}(C^n_{N}-1)/2$ pairs of $(\vec{\chi}_n,\vec{\chi}'_n)$ with $\vec{\chi}_n$ and $\vec{\chi}'_n$ distinct (regardless of the order of the two), there are actually $C^n_{N}N(N-n)/2$ such pairs in which $\vec{\chi}_n$ and $\vec{\chi}'_n$ share exactly $n-1$ elements. For a given pair $(\vec{\eta}_n,\vec{\eta}'_n)$ having this property, assuming that it is $\eta_\beta$ ($\eta'_{\beta'}$) that does not appear in the given set $\{\vec{\eta}'_n\}$ ($\{\vec{\eta}_n\}$), then from Eq.~(\ref{s_ijllLP}) we have
\begin{eqnarray}\label{s_ijllLPn-1}
&&\bar{F}^{\mathrm{( p)}}_{l,l;\vec{\eta}_{n},\vec{\eta}'_{n}}=(-1)^{\beta+\beta'+1}  U^{(\sigma_{n})*}_{\eta_\beta,l}  U^{(\sigma_{n}) }_{\eta'_{\beta'},l}.
\end{eqnarray}
\par The case of $l=l'$ considered above is interesting since $\bar{F}^{\mathrm{( p)}}_{l,l;\vec{\chi}_{n},\vec{\chi}'_{n}}$  can give the matrix elements of operators such as $S^z_j$, $S^z_jS^z_{j+1}$, etc., which are relevant to the central spin model, the Heisenberg spin chain, and so on. Actually, from the relation $S^-_lS^+_l=\frac{1}{2}-S^z_l$, we can define
\begin{eqnarray}\label{SOMsz_def}
G^{\mathrm{( p)}}_{l;\vec{\chi}_{n},\vec{\chi}'_{n}}&\equiv&\langle \vec{\chi}^{\mathrm{( p)}}_{n}|S^z_l|\vec{\chi}'^{\mathrm{( p)}}_{n}\rangle\nonumber\\
&=&\frac{1}{2}\delta_{\vec{\chi}_n,\vec{\chi}'_n}-\bar{F}^{\mathrm{( p)}}_{l,l;\vec{\chi}_{n},\vec{\chi}'_{n}},
\end{eqnarray}
which also vanishes for $\{\vec{\chi}_{n}\}$ and $\{\vec{\chi}'_{n}\}$ sharing fewer than $n-1$ elements, as well as
\begin{eqnarray}\label{SOMszsz_def}
\bar{G}^{\mathrm{( p)}}_{l,l';\vec{\chi}_{n},\vec{\chi}'_{n}}&\equiv&\langle \vec{\chi}^{\mathrm{( p)}}_{n}|S^z_lS^z_{l'}|\vec{\chi}'^{\mathrm{( p)}}_{n}\rangle\nonumber\\
&=&\sum_{\vec{\eta}_n}G^{\mathrm{( p)}}_{l;\vec{\chi}_{n},\vec{\eta}_{n}}G^{\mathrm{( p)}}_{l';\vec{\eta}_{n},\vec{\chi}'_{n}},
\end{eqnarray}
where the set of summation indices $\{\vec{\eta}_n\}$ in the above equation has $n$ or $n-1$ common elements with both $\vec{\chi}_n$ and $\vec{\chi}'_n$, so that the two sets $\{\vec{\chi}_n\}$ and $\{\vec{\chi}'_n\}$ have at most two distinct elements. One can also show that there are $C^n_{N}C^2_NC^2_{N-n}/2$ such pairs of $(\vec{\chi}_n,\vec{\chi}'_n)$ in which the two have exactly $n-2$ common elements. When $N$ and $n$ are relatively large, we have $C^n_N\gg C^2_NC^2_{N-n}$, and hence the total numbers of matrix elements to be considered can be greatly reduced.
\par Combining Eq.~(\ref{s_ijllLP}) with Eq.~(\ref{SOMsz_def}), the collective matrix elements of $\sum^N_{j=1}g'_jS^z_j$ for some distribution $\{g'_j\}$ can be calculated as
\begin{eqnarray}\label{SOMsz_def_g_j}
&&G^{\mathrm{( p)}}_{\vec{\chi}_{n},\vec{\chi}'_{n}}(\{g'_j\})\equiv\langle \vec{\chi}^{\mathrm{( p)}}_{n}|\sum^N_{j=1}g'_jS^z_j|\vec{\chi}'^{\mathrm{( p)}}_{n}\rangle\nonumber\\
&=&-\frac{1}{2}\delta_{\vec{\chi}_n,\vec{\chi}'_n}\left(\sum^N_{j=1}g'_j\right) +\sum^n_{\alpha=1} \sum^n_{\alpha'=1}(-1)^{\alpha+\alpha'} \nonumber\\
&&\left(\sum^N_{j=1}g'_j U^{(\sigma_{n})*}_{\chi_\alpha,j}  U^{(\sigma_{n}) }_{\chi'_{\alpha'},j}\right) \delta_{\vec{\chi}^{(\alpha)}_n,\vec{\chi}'^{(\alpha')}_n},
\end{eqnarray}
It can be shown that $G^{\mathrm{( p)}}_{\vec{\chi}_{n},\vec{\chi}'_{n}}(\{g'_j\})$ can also be written as~\cite{PRB2016}
\begin{eqnarray}\label{Ggj}
&&G^{\mathrm{( p)}}_{\vec{\chi}_{n},\vec{\chi}'_{n}}(\{g'_j\})=-\frac{1}{2}\delta_{\vec{\chi}_n,\vec{\chi}'_n}\left(\sum^N_{j=1}g'_j\right)\nonumber\\
&&+\sum_{\vec{j}_n}\mathcal{S}_{\vec{\chi}'_n,\vec{j}_n}\mathcal{S}^*_{\vec{\chi}_n,\vec{j}_n}\left(\sum^n_{l=1}g'_{j_l}\right).
\end{eqnarray}
\par For a uniform distribution $g'_j=g'$, it can be seen from either Eq.~(\ref{SOMsz_def_g_j}) or Eq.~(\ref{Ggj}) that
\begin{eqnarray}\label{Ggg1}
G^{\mathrm{( p)}}_{\vec{\chi}_{n},\vec{\chi}'_{n}}(g')&=&g'\left(n-\frac{N}{2}\right)\delta_{\vec{\chi}'_n,\vec{\chi}_n}.
\end{eqnarray}
\par In the atomic limit, the transformation matrix $U^{(\sigma)}$ reduces to the identity matrix and the eigenbasis reduces to the Ising configurations in real space. It is shown in Appendix~\ref{AppD} that the matrix representation of the real-space matrix element $\bar{F}^{\mathrm{( p)}}_{l,l';\vec{j}_{n},\vec{j}'_{n}}$ given by Eq.~(\ref{s_ij_1}) survives only for two Ising configurations satisfying $\vec{j}^{(m)}_n=\vec{j}'^{(m')}_n$ for some $m$ and $m'$, where $l=j_m\notin\{\vec{j}'_n\}$ and $l'=j'_{m'}\notin\{\vec{j}_n\}$ are the two distinct elements that are not shared by the two sets. This is consistent with the definition $\bar{F}^{\mathrm{( p)}}_{l,l';\vec{j}_{n},\vec{j}'_{n}}=\langle \vec{j}_n|S^+_{l}S^-_{l'}|\vec{j}'_n\rangle$ that is often used in conventional diagonalization of spin models, where $l$ and $l'$ correspond to the two sites that are connected by $S^+_lS^-_{l'}$. Thus, Eq.~(\ref{s_ij_1}) also provides a \emph{compact way} to calculate real-space matrix elements of the $XY$-type spin-spin interaction in interacting spin chains. For the Heisenberg model described by $H_{\rm{Heisenberg}}=H_{\rm{XX}}+H_{\rm{Ising}}$, where $H_{\rm{Ising}}=\sum^N_{j=1}J'_{j} S^z_jS^z_{j+1}$, it is interesting to note that $H_{\rm{XX}}$ is diagonal in the eigenbasis $\{|\vec{\chi}_n\rangle\}$, while $H_{\rm{Ising}}$ is diagonal in the real basis $\{|\vec{j}_n\rangle\}$.


\section{Applications}
\par In this section, we apply the results obtained in the preceding section to two physical systems, namely the nonlinear optical response of a one-dimensional molecular aggregate studied in Ref.~\cite{Spano1991}, and the real-time dynamics of an interacting Dicke model. Whereas the latter problem can in principle also be dealt with by other numerical methods, the nonlinear response of molecular aggregates requires essentially the information of matrix elements of the transition dipole operator in the \emph{energy basis} of the aggregates.
\subsection{Nonlinear optical response of one-dimensional molecular aggregates}
As mentioned in the Introduction, the one-dimensional XX spin chain described by Eq.~(\ref{XX}) can model a linear molecular aggregate consisting of an array of coupled two-level molecules, with $J_j$ and $h_j$ being the nearest-neighbor dipole-dipole coupling and the optical two-level transition frequency of the $j$th molecule, respectively. Taking advantage of the fact that the fundamental electronic excitations in such a system are in fact fermions, Spano proposed a simplified way to calculate the third-order hyperpolarizability for an aggregate with site disorder~\cite{Spano1991}. The third-order response requires knowledge of the one- and two-exciton eigenstates and eigenenergies, for which the obtained determinant representation of the transition dipole matrix elements can be directly used.
\par To calculate the nonlinear optical response, the matrix elements of the transition dipole operator (with $\mu$ the transition dipole moment between the ground and excited state of the two-level molecule)
\begin{eqnarray}\label{TDO}
\hat{\mu}=\mu\sum^N_{j=1}(S^\dag_j+S^-_j)
\end{eqnarray}
are needed, where we assumed that the dimension of the aggregate is small enough compared with the optical wavelength. In particular, the third-order aggregate hyperpolarizability $\gamma(-\omega;\omega,\omega,-\omega)$ is related to the matrix elements~\cite{Spano1991},
\begin{eqnarray}\label{mu01}
\mu_{0,\eta}&\equiv&\langle 0|\hat{\mu}\xi^\dag_\eta|0\rangle,~\eta=1,2,\cdots,N\nonumber\\
\mu_{\eta,\chi_1\chi_2}&\equiv&\langle 0|\xi_{\eta}\hat{\mu}\xi^\dag_{\chi_1}\xi^\dag_{\chi_2}|0\rangle,~1\leq\chi_1<\chi_2\leq N\nonumber
\end{eqnarray}
which connect the vacuum state $|0\rangle$ to the $N$ one-exciton states $|\eta\rangle$, and connect the latter to the $C^2_N$ two-exciton states $|\chi_1,\chi_2\rangle$, respectively. We recognize that $\mu_{0,\eta}$ ($\mu_{\eta,\chi_1\chi_2}$) is just the collective SOME defined in Eq.~(\ref{SOM_gj_def}) with $n=0$ ($n=1$),
\begin{eqnarray}\label{mu011}
\mu_{0,\eta}&=&\mu F_{\eta,0}(\{g_j=1\}),\\
\mu_{\eta,\chi_1\chi_2}&=&\mu F_{\chi_1\chi_2,\eta}(\{g_j=1\}).
\end{eqnarray}
\par For a homogeneous molecular chain with periodic boundary conditions, these matrix elements are given by the factorized expression in Eq.~(\ref{DI}) (for $N=$even molecules):
\begin{eqnarray}\label{mu012}
&&\mu^{\mathrm{(p)}}_{0,\eta}=\mu \sqrt{N}\delta_{\eta,\frac{N}{2}+1},\\
&&\mu^{\mathrm{(p)}}_{\eta,\chi_1\chi_2}=\mu \frac{2}{\sqrt{N}}\delta(K^{\mathrm{(+1)}}_{\chi_1}+K^{\mathrm{(+1)}}_{\chi_2}-K^{\mathrm{(-1)}}_{\eta})\cdot\nonumber\\
&&\frac{e^{iK^{\mathrm{(+1)}}_{\chi_2}}-e^{iK^{\mathrm{(+1)}}_{\chi_1}}}{[1-e^{-i(K^{\mathrm{(+1)}}_{\chi_1}-K^{\mathrm{(-1)}}_{\eta})}][1-e^{-i(K^{\mathrm{(+1)}}_{\chi_2}-K^{\mathrm{(-1)}}_{\eta})}]}.
\end{eqnarray}
We see that $\mu^{\mathrm{(p)}}_{0,\eta}$ vanishes unless $K^{(-1)}_\eta=0$, and $\mu^{\mathrm{(p)}}_{\eta,\chi_1\chi_2}$ is nonzero only if $K^{\mathrm{(+1)}}_{\chi_1}+K^{\mathrm{(+1)}}_{\chi_2}-K^{\mathrm{(-1)}}_{\eta}=0$ or $\pm 2\pi$, implying the momentum conservation of excitons in the optical response.
\par For a homogeneous molecular chain with free ends studied in Ref.~\cite{Spano1991}, the corresponding $\mu^{\mathrm{(o)}}_{0,\eta}$ and $\mu^{\mathrm{(o)}}_{\eta,\chi_1\chi_2}$ can also be calculated analytically from Eqs.~(\ref{A_tilde}) and (\ref{SOM_cc_open_1}),
\begin{eqnarray}\label{mu022}
&&\mu^{\mathrm{(o)}}_{0,\eta}=\mu \sqrt{\frac{2}{N+2}}\frac{1-(-1)^\eta}{2}\cot\frac{K_\eta}{2},
\end{eqnarray}
\begin{eqnarray}\label{mu023}
&&\mu^{\mathrm{(o)}}_{\eta,\chi_1\chi_2}\nonumber\\
&=&\mu \sqrt{\frac{1}{2(N+1)}}\cot\frac{K_{\chi_1}}{2}[-(\delta_{\chi_1+\eta,\chi_2}+\delta_{\chi_1+\chi_2,\eta})\nonumber\\
&&+(\delta_{\chi_1,\chi_2+\eta} +\delta_{\chi_1+\chi_2+\eta,2(N+1) })]-(\chi_1\leftrightarrow \chi_2)
\end{eqnarray}
for $\eta\neq\chi_1$ and $\eta\neq\chi_2$. For $\eta=\chi_1$(and hence $\eta\neq\chi_2$ since $\chi_1\neq \chi_2$), we have
\begin{eqnarray}\label{mu024}
&&\mu^{\mathrm{(o)}}_{\chi_1,\chi_1\chi_2}\nonumber\\
&=&-\mu \sqrt{\frac{1}{2(N+1)}} [1-(-1)^{\chi_2-1}]\cot\frac{K_{\chi_2}}{2}\nonumber\\
&& - \mu \sqrt{\frac{1}{2(N+1)}}\left(\cot\frac{K_{\chi_1} }{2} +\cot K_{\chi_1}\right)\cdot\nonumber\\
&&(\delta_{2\chi_1,\chi_2 } -\delta_{\chi_2+2\chi_1,2(N+1) }).
\end{eqnarray}
Similarly, when $\eta=\chi_2$ (and hence $\eta\neq \chi_1$), $\mu^{\mathrm{(o)}}_{\chi_2,\chi_1\chi_2}$ can be obtained from $\mu^{\mathrm{(o)}}_{\chi_1,\chi_1\chi_2}$ by swapping $\chi_1$ and $\chi_2$ and noting that $\mu^{\mathrm{(o)}}_{\chi_2,\chi_1\chi_2}=-\mu^{\mathrm{(o)}}_{\chi_2,\chi_2\chi_1}$. We see from the above expressions that $\mu^{\mathrm{(o)}}_{0,\eta}$ is nonzero only if $\eta$ is odd~\cite{Spano1991}, while $\mu^{\mathrm{(o)}}_{\eta,\chi_1\chi_2}$ vanishes unless $\eta$ and $\chi_1+\chi_2$ have the same parity~\cite{mu12}.
\begin{figure}
\includegraphics[width=.50\textwidth]{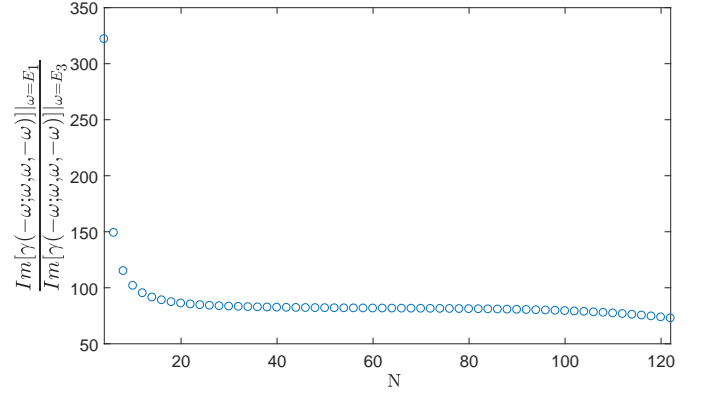}
\caption{The ratio of the excitonic absorption peaks occurring at the first peak $\omega=E_1$ to that at the second peak $\omega=E_3$, i.e., $\Im[\gamma(-\omega;\omega,\omega,-\omega)]|_{E_1}/\Im[\gamma(-\omega;\omega,\omega,-\omega)]|_{E_3}$, for homogeneous aggregates of different sizes [see Ref.~\cite{Spano1991} for an explicit expression of the third-order aggregate hyperpolarizability $\gamma(-\omega;\omega,\omega,-\omega)$]. Open boundary conditions are assumed for the molecular aggregates. Other parameters are set the same as those in Ref.~\cite{Spano1991}.}
\label{Fig0}
\end{figure}
\par For a homogeneous aggregate with free ends, the saturated excitonic absorption spectra show peaks at $\omega=E_{1},~E_3,~E_5,\cdots$~\cite{Spano1991}. Figure~\ref{Fig0} shows the evolution of the ratio of the imaginary parts of the hyperpolarizability $\gamma(-\omega;\omega,\omega,-\omega)$ (see Ref.~\cite{Spano1991} for an explicit expression) at the first peak $\omega=E_1$ to that at the second peak $\omega=E_3$ with the number of molecules $N$ in the aggregate. It can be seen that the ratio $\Im[\gamma(-E_1;E_1,E_1,-E_1)]/\Im[\gamma(-E_3;E_3,E_3,-E_3)]$ drops quickly as $N$ is increased from small $N$, and it keeps decreasing more moderately as $N$ increases further up to several tens, which is consistent with Ref.~\cite{Spano1991}.
\par There may be cases in which the aggregates sizes are comparable with the optical wavelength~\cite{Spano-Mu}. As a result, the transition dipole operator can generally no longer be written as a collective form given by Eq.~(\ref{TDO}), but rather as $\hat{\mu}=\mu\sum^N_{j=1}(e^{i\mathbf{k}\cdot \mathbf{r}_j}S^\dag_j+\mathrm{H.c.})$, where $\mathbf{k}$ and $\mathbf{r}_j$ are the wave vector of the light and the position of the $j$th molecule, respectively. In turn, the matrix elements for the transition dipole operator do not admit closed forms such as Eqs.~(\ref{mu01})-(\ref{mu024}) and must be evaluated through Eq.～(\ref{SOM_gj_def}) by calculating all the $NC^n_NC^{n+1}_N$ individual matrix elements $F_{j;\vec{\eta}_{n+1},\vec{\chi}_n}$. Thanks to the determinant form of $F_{j;\vec{\eta}_{n+1},\vec{\chi}_n}$ obtained in Eq.~(\ref{SOM_cc_final}), the lengthy sums over the site indices in Eq.~(\ref{f_SS})~\cite{Spano1991} are avoided. We note that if periodic boundary conditions are imposed on the molecular aggregates, the transition dipole matrix elements can be obtained most simply through the factorized expression given by Eq.~(\ref{CME_final2}). The advantages of the present method become more apparent when higher-order nonlinear optical properties are involved~\cite{fifth}, for which multiexciton states with more than two excitons need to be taken into account. The formalism developed above provides a convenient method to calculate these higher-order nonlinear optical responses.
\subsection{The interacting Dicke model and its dynamics}
\par As the second application of the developed formalism for the SOMEs, we now turn to the study of real-time dynamics of the interacting Dicke model. Specifically, we consider a model consisting of a periodic XX spin chain coupled to a single bosonic mode:
\begin{eqnarray}\label{H_int}
H_{\mathrm{int}}&=&H_{\mathrm{int},0}+H_{\mathrm{int},1},\nonumber\\
H_{\mathrm{int},0}&=&\sum^N_{j=1}\omega_j\left(S^z_j+\frac{1}{2}\right)+J\sum^{N}_{j=1}(S^x_jS^x_{j+1}+S^y_jS^y_{j+1})\nonumber\\
&&+\omega a^\dag a,\nonumber\\
H_{\mathrm{int},1}&=&\sum^N_{j=1}g_j(S^+_ja+S^-_ja^\dag),
\end{eqnarray}
where $a$ is the boson annihilation operator for the single-mode photon with frequency $\omega$. In the Frenkel-exciton model description of molecular aggregates located in a single-mode cavity, $\{\omega_j\}$ and $\{g_j\}$ are the (inhomogeneous) excitonic excitation energies and the exciton-photon coupling constants, respectively. $J$ is the uniform nearest-neighbor exciton coupling between adjacent monomers arranged in a line~\cite{Hanamura,2016LPP}. It is easily seen that $H_{\mathrm{int}}$ conserves the total number of excitations $M=a^\dag a+\sum_j (S^z_j+1/2)$, implying that $H_{\mathrm{int}}$ can be diagonalized in subspaces with fixed $M$'s.
\par The usual inhomogeneous Dicke model $H_{\mathrm{Dicke}}$ that describes a set of $N$ two-level atoms interacting with a single-photon mode~\cite{Loss2009,Loss2010} can be obtained by setting $J=0$ in $H_{\mathrm{int}}$. In the special case of a uniform light-atom interaction $g_j=g$, $H_{\mathrm{Dicke}}(g)$ is integrable and can be solved by using the Bethe ansatz~\cite{Fari2012,NPB}. The dynamics of the inhomogeneous Dicke model has been studied in detail in Refs.~\cite{Loss2009,Loss2010,Fari2012}. Note that $H_{\mathrm{int}}$ is no longer integrable, and hence the Bethe ansatz ceases to be applicable.
\par To obtain a universal short-dynamics for different numbers of spins, we define the collective Rabi frequency~\cite{PRB2016,2016LPP}
\begin{eqnarray}\label{Rabi}
g_R=\sqrt{\sum^N_{j=1}g^2_j},
\end{eqnarray}
which will be used as an energy unit below. The Hilbert space in the $M$-sector is spanned by the $D_{N,M}=\sum^{\min\{M,N\}}_{m=0}C^m_N$ basis states $\{|\vec{\eta}_m;M-m\rangle\}$ in which $H_{\mathrm{int},0}$ is diagonal, where $m$ counts the total number of excitations in the XX chain. So any state in the $M$-sector can be written
\begin{eqnarray}\label{eig_Dicke}
|\psi_M\rangle=\sum^{\min\{M,N\}}_{m=0}\sum_{\vec{\eta}_m}A^{(m)}_{\vec{\eta}_m}|\vec{\eta}_m;M-m\rangle.
\end{eqnarray}
The matrix element of $H_{\mathrm{int}}$ between any two basis states $|\vec{\eta}_m;M-m\rangle$ and $|\vec{\eta}'_{m'};M-m'\rangle$ reads
\begin{eqnarray}\label{Hme}
&&\langle\vec{\eta}'_{m'};M-m'|H_{\mathrm{int}}|\vec{\eta}_m;M-m\rangle\nonumber\\
&=&\delta_{m,m'}\prod^m_{l=1}\delta_{\eta'_l,\eta_l}[\mathcal{E}_{\vec{\eta}_m}+\omega(M-m)]\nonumber\\
&&+\delta_{m',m+1}F^*_{\vec{\eta}'_{m+1},\vec{\eta}_m}(\{g_j\})\sqrt{M-m}\nonumber\\
&&+\delta_{m',m-1}F_{\vec{\eta}_{m},\vec{\eta}'_{m-1}}(\{g_j\})\sqrt{M-m+1},
\end{eqnarray}
where $\mathcal{E}_{\vec{\eta}_m}=\sum^m_{l=1}E_{\eta_l}$ is the total energy of the $m$ fermions occupying the set of modes $\{\vec{\eta}_m\}$.
\par In the following numerical simulation, we use Eqs.~(\ref{SOM_gj_def}) and (\ref{SOM_cc_final}) to compute the collective matrix elements $F_{\vec{\eta}_{n+1},\vec{\eta}_n}$ appearing in Eq.~(\ref{Hme}). Once the block Hamiltonian in the $M$-sector is constructed, the time-evolved state $|\psi(t)\rangle=e^{-iHt}|\psi_0\rangle$ from an initial state $|\psi_0\rangle$ is then calculated by numerically integrating the Schr\"odinger equation $i\partial_t|\psi(t)\rangle=H|\psi(t)\rangle$. We also consider numbers of excitations no larger than the total number of spins, i.e., $M\leq N$. We emphasize that the dynamics of the same model can in principle also be treated in the real-space basis of the XX chain. However, Eq.~(\ref{Hme}) offers us a compact expression for evaluating the matrix elements of the Hamiltonian provided the SOMEs $F_{\vec{\eta}_{n+1},\vec{\eta}'_{n}}(\{g_j\})$ are obtained.
\subsubsection{Noninteracting chain, homogeneous coupling}
\par Using a combination of mean-field analysis and algebraic Bethe ansatz, the authors of Ref.~\cite{Fari2012} studied the decay of the bosonic occupation number $N_a(t)=\langle\psi(t)|a^\dag a|\psi(t)\rangle$ in the integrable model described by $H_{\mathrm{Dicke}}(g)$, with uniformly distributed spin excitation energies $\omega_j=(j-1)\frac{\Delta}{N-1}$ between zero and the bandwidth $\Delta$. It was pointed out in Ref.~\cite{Fari2012} that neither the mean-field approach nor the Bethe ansatz solution, which requires a truncation of the Hilbert space, can capture the real-time dynamics of the system accurately in the intermediate-coupling regime with $\Delta/N<g_R=g\sqrt{N}<\Delta$ due to the significant mixture between the spins and the bosonic mode. Therefore, a full quantum treatment is needed in this regime, which limits the number of spins considered up to only $N\leq16$. For example, for $N=16$ spins and $M=6$ excitations, the dimension of the Hilbert space in this $M$-sector already reaches $D_{16,6}=14893$.

\par
\begin{figure}
\includegraphics[width=.52\textwidth]{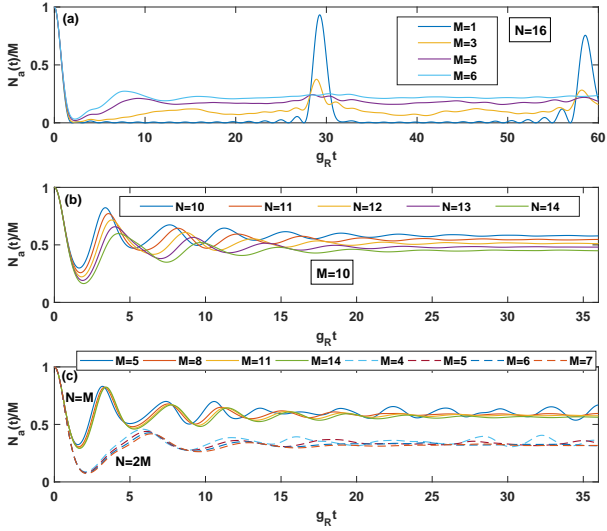}
\caption{Real-time evolution of the reduced bosonic occupation number $N_a(t)/M$  for $J=0$, $\Delta/g_R=10/3$, and $\omega/g_R=5/3$. The initial state is a boson number state $|\psi_0\rangle=|\downarrow\cdots\downarrow;M\rangle$ occupied by $M$ bosons. (a) The number of spins is $N=16$, and the numbers of excitations are $M=1,3,5$, and $6$; (b) The number of excitations is $M=10$, and the numbers of spins are $N=10, 11, 12, 13$, and $14$; (c) The ratio between $N$ and $M$ is fixed: $N/M=1$ (solid curves) and $N/M=2$ (dashed curves).}
\label{Fig1}
\end{figure}

 As in Ref.~\cite{Fari2012}, we are interested in the time evolution of the reduced bosonic occupation number $N_a(t)/M$ starting with the initial state $|\psi_0\rangle=|\downarrow\cdots\downarrow;M\rangle$. Figure~\ref{Fig1} shows $N_a(t)/M$ for different combinations of $(N,M)$. The ratio between the excitonic bandwidth and the collective Rabi frequency is set as a constant $\Delta/g_R=10/3$ for different numbers of spins, so that the system lies in the intermediate-coupling regime. The photon energy is always set to be half of the bandwidth, i.e., $\omega/\Delta=0.5$. In Fig.~1(a), we show the evolution of $N_a(t)/M$ for $N=16$ spins. For small numbers of excitations with $M=1$ and $3$, we observe revivals of $N_a(t)/M$ at later times $g_Rt\approx 30$ and $\approx60$, which are mainly due to the finite-size effect for small $M$. As $M$ increases, the revival behavior disappears gradually, and plateaus are developed at intermediate and long times due to a rapid increase of the dimension of the Hilbert space. In addition, the plateau value tends to increase with increasing $M$, which is consistent with the observation that an increase of $M$ tends to suppress the decay of $N_a(t)/M$ at short times~\cite{Fari2012}. However, our results go beyond the short-time dynamics obtained in Ref.~\cite{Fari2012} to reach the steady long-time regime. Furthermore, even though in the framework of Bethe ansatz solutions one can perform a full quantum calculation as well, the nasty double sum over all the eigenstates cannot be avoided~\cite{Fari2012}.
\par Figure 1(b) shows the evolution of $N_a(t)/M$ for a fixed excitation number $M=10$, but with different numbers of spins. In contrast to the case of varying $M$ and keeping $N$ fixed, increasing $N$ with $M$ fixed can actually pull the profile of $N_a(t)/M$ down, implying that it might be the ratio $M/N$ that qualitatively determines the overall profile of $N_a(t)/M$. This is confirmed in Fig.~1(c) for $M/N=1/2$ and $1$. In both cases, some slight oscillations appear in the plateau regime for small $M$. However, the curves become closer to each other as $M$ is increased, and they are expected to converge to a single curve in the limit $M,N\to\infty$.
\subsubsection{Homogeneous chain, inhomogeneous coupling}
\par We now go beyond the atomic limit to include finite dipole-dipole interaction between nearest-neighbor monomers. For simplicity, we impose periodic boundary conditions on the chain and assume uniform on-site energies for the monomers, i.e., $\omega_j=\omega_A,~\forall j$. Thus, the noninteracting $H_{\mathrm{int},0}$ becomes
\begin{eqnarray}
H_{\mathrm{int},0}&=&\omega M+H_{\mathrm{PBC}}(-\Delta_{AC},J),
\end{eqnarray}
with single-particle dispersion $E_{\eta,\sigma}=\Delta_{AC}+2J\cos K^{(\sigma)}_\eta$ in the sector with fermion parity $\sigma$, where $\Delta_{AC}=\omega_A-\omega$ is the exciton-cavity detuning. For $J<0$ ($J>0$), the Hamiltonian $H_{\mathrm{int},0}$ describes a linear J-aggregate (H- aggregate)~\cite{Spanoreview}. With the bosonic mode describing a single-mode cavity coupled to the aggregate, the exciton-cavity coupling is assumed to be of the form
\begin{eqnarray}\label{gj}
g_j=g_d\sin\frac{\pi j}{2N},
\end{eqnarray}
\begin{figure}
\includegraphics[width=.52\textwidth]{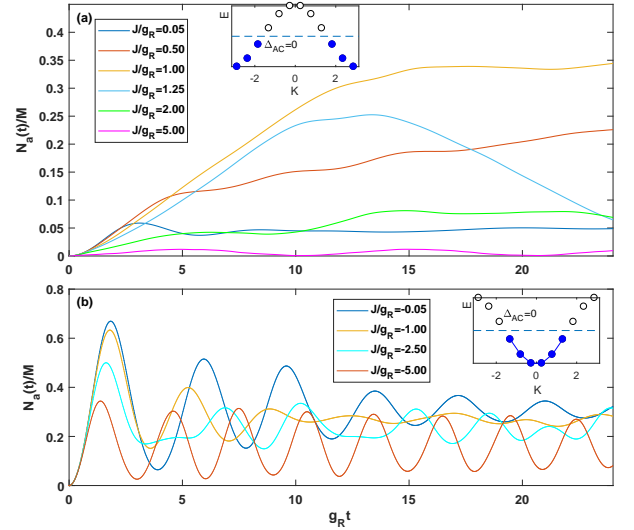}
\caption{Real-time evolution of the reduced photon number $N_a(t)/M$ in the resonant case $\omega=\omega_A$ for $N=12$ monomers. (a) $J>0$, (b) $J<0$. The initial state is chosen as the ground state of the excitons with $M=6$ excitations, where the occupied excitonic modes are indicated as solid blue circles in the corresponding insets.}
\label{Fig2}
\end{figure}
where $g_d$ measures the coupling strength and is related to the dipole moment of the exciton, and the sinusoidal part is due to different positions of the monomers in the cavity~\cite{PRA2009}. The inhomogeneous coupling constants $\{g_j\}$ are thus distributed nonuniformly between $g_1=g_d\sin\frac{\pi}{2N}$ and $g_N=g_d$. For the resonant case with $\Delta_{AC}=0$, the ratio between the collective Rabi frequency and the exciton bandwidth $g_R/2|J|$ provides a measure of the exciton-cavity coupling strength.
\par In this subsection, we consider initial states with all excess energy contained in the excitonic part. In particular, we consider the following initial state in the $M$ sector:
\begin{eqnarray}\label{Hkmode}
|\psi_0\rangle=|G\rangle_M\otimes|0\rangle_{\rm{c}},
\end{eqnarray}
which is a product state of the excitonic ground state $|G\rangle_M$ with $M$ excitons and the vacuum state $|0\rangle_{\rm{c}}$ of the cavity photons. Depending on the sign of $J$, the ground state is filled by $M$ fermions with their wavenumbers distributed at the middle (edges) of the Brillouin zone $(-\pi,\pi)$ for $J<0$ ($J>0$). In Fig.~\ref{Fig2}, we present the dynamics of $N_a(t)/M$ for a molecular chain with $N=12$ monomers and $M=6$ excitations, so that the ground state of the molecular chain is $|G\rangle_6=\xi^\dag_{4}\xi^\dag_{5}\cdots\xi^\dag_{9}|0\rangle$ and $|G\rangle_6=\xi^\dag_{1}\xi^\dag_{2}\xi^\dag_{3}\xi^\dag_{10}\xi^\dag_{11}\xi^\dag_{12}|0\rangle$ for $J<0$ and $J>0$, respectively (Fig.~\ref{Fig2}, insets).
\par For $J>0$, we observe a nonmonotonic dependence of the photon generation on the exciton coupling $J$ [Fig.~\ref{Fig2}(a)]. In the strong exciton-cavity coupling regime with $J/g_R\ll1$, the photon generation from the half-filled exciton ground state is suppressed since the exciton-cavity coupling is strong enough to excite $|G\rangle_6$ into a large number of excitonic excited states due to the narrow exciton band width. As the exciton coupling $J$ increases, the energy differences between different eigenenergies of the molecular chain also increase. If the exciton-cavity coupling can efficiently excite higher occupied excitonic modes into the cavity mode, and at the same time excitations to the unoccupied excitonic modes are effectively suppressed, then the photon generation is enhanced mostly. When $J$ is increased further, so that even the smallest excitation energy exceeds the largest exciton-cavity coupling $g_N$, then the photon generation is again suppressed.
\par For $J<0$, the excitonic ground state $|G\rangle_6$ has a different nature from that of $J>0$, which leads to an oscillatory decay of $N_a(t)/M$. Though in this case there is no clear indicator of the photon generation behavior, it seems that the mean value of $N_a(t)/M$ shows a similar nonmonotonic dependence on $|J|$. For large enough $|J|$, we again observe a suppression of the photon generation. These behaviors of the photon generation are similar to the decoherence properties previously observed in an interacting central spin model~\cite{PRB2016}.

\section{Conclusions}
\label{sec-final}
\par In this work, we derived compact expressions for the spin-operator matrix elements (SOMEs) of spin operators $S^-_j$ and $S^-_jS^+_{j'}$ between too relevant eigenstates of an inhomogeneous periodic/open XX spin chain. Using a fermionic approach that has been applied to the calculation of SOMEs in the quantum Ising model, we show that these matrix elements can simply be expressed as determinants of some square matrices that involve the coefficients of the canonical transformations diagonlizing the chain. For a homogeneous XX chain with periodic boundary conditions, the SOME of $S^-_j$ happens to be proportional to a variant of the Cauchy determinant that can be evaluated analytically, which recovers a known but unproven result discovered in Ref.~\cite{Hanamura}. Using the results for the SOMEs of $S^-_jS^+_{j'}$ in the case of $j=j'$, we also derived useful expressions for the SOMEs of $S^z_{l}$ and $S^z_lS^z_{l'}$, which are relevant to the central spin model and the Heisenberg spin chain, etc..
\par We then applied the obtained formalism to the study of third-order optical response of molecular aggregates with free ends. Since the transition dipole matrix elements between eigenstates of an aggregate are essential for the calculation of the nonlinear optical responses, our results thus provide a suitable framework for this purpose, especially when the molecule sizes are comparable with the optical wavelength, or when higher-order responses need to be considered. We next studied the real-dynamics of an interacting Dicke model that describes a set of interacting spins 1/2 coupled to a single bosonic model. In the noninteracting case, we obtain results that are consistent with the literature, and we find that it is the ratio between the number of excitations and the total number of spins that determines the overall profile of the reduce bosonic occupation number. In the interacting case that is relevant to a one-dimensional molecular chain located in a single-mode cavity, we find that the exciton coupling between nearest-neighboring monomers has a significant effect on the photon generation from a half-filled exctonic ground state. We believe the results obtained in this work can find useful applications in a variety of composite ``system-environment" systems, e.g., low-dimensional molecular aggregates, optical emitters in photocell systems, interacting central spin models, and so on.

\noindent{\bf Acknowledgements:}

We are grateful to Hosho Katsura for insightful suggestions and useful discussions. This work was supported by the NSFC under Grant No. 11705007 and partially by a startup fund from the Beijing Institute of Technology.

\appendix
\section{Derivation of Eq.~(\ref{f_SS})}\label{AppA}
From Eq.~(\ref{jnkn}) and Eq.~(\ref{SOM_gj_def}), $F^{\mathrm{( p)}}_{\vec{\eta}_{n+1},\vec{\chi}_n}(\{g_j\})$ can be written as
\begin{eqnarray}\label{SOM_gj_def_App}
&&F^{\mathrm{( p)}}_{\vec{\eta}_{n+1},\vec{\chi}_n}(\{g_j\})\nonumber\\
&=&\langle \vec{\chi}^{\mathrm{( p)}}_n|\sum_{\vec{j}_{n+1}}\sum^N_{j=1}g_j\mathcal{S}_{\vec{\eta}_{n+1};\vec{j}_{n+1}}T_jc_j|\vec{j}_{n+1}\rangle.
\end{eqnarray}
Since $c_j|0\rangle=0$, the summation index $j$ in the above equation should be chosen from the set $\{\vec{j}_{n+1}\}=\{j_1,j_2,\cdots,j_{n+1}\}$, yielding
 \begin{eqnarray}\label{SOM_cc1}
&&F^{\mathrm{( p)}}_{\vec{\eta}_{n+1},\vec{\chi}_n}(\{g_j\})\nonumber\\
&=&  \langle \vec{\chi}^{\mathrm{( p)}}_n|\sum_{\vec{j}_{n+1}}\sum^{n+1}_{l=1}g_{j_l}\mathcal{S}_{\vec{\eta}_{n+1};\vec{j}_{n+1}}T_{j_l}c_{j_l}\prod^{n+1}_{m=1}c^\dag_{j_m}|0\rangle.
\end{eqnarray}
By using Eq.~(\ref{TCT}) to move $T_{j_{l}}c_{j_l}$ across the creations operators $c^\dag_{j_1},\cdots,c^\dag_{j_{l-1}}$, and noting that $T_{j_l}|0\rangle=|0\rangle$, we have
 \begin{eqnarray}\label{SOM_cc_App}
&&F^{\mathrm{( p)}}_{\vec{\eta}_{n+1},\vec{\chi}_n}(\{g_j\})\nonumber\\
&=&\langle \vec{\chi}^{\mathrm{( p)}}_n|\sum_{\vec{j}_{n+1}}\sum^{n+1}_{l=1}g_{j_l}\mathcal{S}_{\vec{\eta}_{n+1};\vec{j}_{n+1}}\prod^{l-1}_{m=1}c^\dag_{j_m}\prod^{n+1}_{m=l+1}c^\dag_{j_m}|0\rangle\nonumber\\
&=&\langle \vec{\chi}^{\mathrm{( p)}}_n|\sum_{\vec{j}_{n+1}}\sum^{n+1}_{l=1}g_{j_l}\mathcal{S}_{\vec{\eta}_{n+1};\vec{j}_{n+1}}\sum_{\vec{\chi}'_n}\mathcal{S}^*_{\vec{\chi}'_n;\vec{j}^{(l)}_{n+1}}|\vec{\chi}'^{\mathrm{(p)}}_n\rangle\nonumber\\
&=&\sum_{\vec{j}_{n+1}}\sum^{n+1}_{l=1}g_{j_l}\mathcal{S}_{\vec{\eta}_{n+1};\vec{j}_{n+1}} \mathcal{S}^*_{\vec{\chi}_n;\vec{j}^{(l)}_{n+1}},
\end{eqnarray}
where we have used the inverse transformation of Eq.~(\ref{jnkn}) in the second to last line.
\section{Derivation of Eq.~(\ref{s_ij_1}) using the fermionic approach}\label{AppB}
In this appendix, we closely follow the method used in Sec.~\ref{sectionIIA} of deriving $F^{\mathrm{( p)}}_{j;\vec{\eta}_{n+1},\vec{\chi}_{n}}$ to derive a determinant representation for $\bar{F}^{\mathrm{( p)}}_{l,l';\vec{\chi}_{n},\vec{\chi}'_{n}}$.
\par We again switch to the fermion representation by writing
\begin{eqnarray}\label{Fll_fermion}
&&\bar{F}^{\mathrm{( p)}}_{l,l';\vec{\chi}_{n},\vec{\chi}'_{n}}\nonumber\\
&=&\langle0|\xi_{\chi_{n},\sigma_n}\cdots \xi_{\chi_{1},\sigma_n}c_{l'}T_{l'}T_{l}c^\dag_{l} \xi^\dag_{\chi'_1,\sigma_n}\cdots \xi^\dag_{\chi'_{n},\sigma_n}|0\rangle\nonumber\\
&=&\sum^N_{\chi=1}\sum^N_{\chi'=1}U^{(\sigma_n)}_{\chi, l'}U^{(\sigma_n)*}_{\chi', l}\bar{D}^{(l,l')}_{\chi,\chi_1,\cdots,\chi_{n};\chi',\chi'_1,\cdots,\chi'_{n}},
\end{eqnarray}
where
\begin{eqnarray}\label{Dll}
&&\bar{D}^{(l,l')}_{\chi,\chi_1,\cdots,\chi_{n};\chi',\chi'_1,\cdots,\chi'_{n}}\nonumber\\
&\equiv&\langle0|\xi_{\chi_n,\sigma_n}\cdots \xi_{\chi_1,\sigma_n}\xi_{\chi,\sigma_n}T_{l'}T_{l}\xi^\dag_{\chi',\sigma_n} \xi^\dag_{\chi'_1,\sigma_n}\cdots \xi^\dag_{\chi'_{n},\sigma_n}|0\rangle.\nonumber\\
\end{eqnarray}
\par
Inserting the identity $(T_{l'}T_{l})(T_{l'}T_{l})=1$ between $\xi_{\eta_{1},\sigma_n}$ and $\xi_{\chi,\sigma_n}$ in the last equation, we have
\begin{eqnarray}\label{Dbar}
&&\bar{D}^{(l,l')}_{\chi,\chi_1,\cdots,\chi_{n};\chi',\chi'_1,\cdots,\chi'_{n}}\nonumber\\
&=&\sum^N_{n'=1}U^{(\sigma_n)*}_{\chi,n'}\langle0|\xi_{\chi_n,\sigma_n}\cdots \xi_{\chi_1,\sigma_n}(T_{l'}T_{l})(T_{l'}T_{l}c_{n'}T_{l'}T_{l})\nonumber\\
&&\xi^\dag_{\chi',\sigma_n}\xi^\dag_{\chi'_1,\sigma_n}\cdots \xi^\dag_{\chi'_{n},\sigma_n}|0\rangle\nonumber\\
&=&\sum^N_{\rho=1}\left[-\sum^{l_{\max}-1}_{n'=l_{\min}} +\sum^{l_{\min}-1}_{n'=1}+\sum^N_{n'=l_{\max}}\right]U^{(\sigma_n)*}_{\chi,n'}U^{(\sigma_n)}_{\rho,n'}\nonumber\\
&&\langle0|\xi_{\chi_n,\sigma_n}\cdots \xi_{\chi_1,\sigma_n} T_{l'}T_{l}\xi_{\rho,\sigma_n}\xi^\dag_{\chi',\sigma_n}\xi^\dag_{\chi'_1,\sigma_n}\cdots \xi^\dag_{\chi'_{n},\sigma_n}|0\rangle\nonumber\\
&=&\sum^N_{\rho=1}\bar{A}^{(l,l'),(\bar{\sigma}_n)}_{\chi,\rho}\langle0|\xi_{\chi_n,\sigma_n}\cdots \xi_{\chi_1,\sigma_n} T_{l'}T_{l}\xi_{\rho,\sigma_n}\nonumber\\
&&\xi^\dag_{\chi',\sigma_n}\xi^\dag_{\chi'_1,\sigma_n}\cdots \xi^\dag_{\chi'_{n},\sigma_n}|0\rangle,
\end{eqnarray}
where $\bar{A}^{(l,l'),(\sigma)}_{\chi,\rho}$ is given by Eq.~(\ref{sum_AA}). Note that only $\rho=\chi',\chi'_1,\cdots,\chi_{n}$ contributes to the last line of Eq.~(\ref{Dbar}), hence we have
\begin{eqnarray}\label{Dbar_LP}
&&\bar{D}^{(l,l')}_{\chi,\chi_1,\cdots,\chi_{n};\chi',\chi'_1,\cdots,\chi'_{n}}=\bar{A}^{(l,l'),(\bar{\sigma}_n)}_{\chi,\chi'}\bar{D}^{(l,l')}_{\chi_1,\cdots,\chi_{n};\chi'_1,\cdots,\chi'_{n}}\nonumber\\
&&+ \sum^{n}_{m=1}(-1)^m\bar{A}^{(l,l'),(\bar{\sigma}_n)}_{\chi,\chi'_m} \bar{D}^{(l,l')}_{\chi_1,\cdots,\chi_{n};\chi',\cdots,\chi'_{m-1},\chi'_{m+1},\cdots\chi'_{n}}\nonumber\\
&=&\det\left(
       \begin{array}{ccccc}
       \bar{A}^{(l,l'),(\bar{\sigma}_n)}_{\chi,\chi'} & \bar{A}^{(l,l'),(\bar{\sigma}_n)}_{\chi_1,\chi'} & \bar{A}^{(l,l'),(\bar{\sigma}_n)}_{\chi_2,\chi'} & \cdot & \bar{A}^{(l,l'),(\bar{\sigma}_n)}_{\chi_{n},\chi'} \\
        \bar{A}^{(l,l'),(\bar{\sigma}_n)}_{\chi,\chi'_1} & \bar{A}^{(l,l'),(\bar{\sigma}_n)}_{\chi_1,\chi'_1} & \bar{A}^{(l,l'),(\bar{\sigma}_n)}_{\chi_2,\chi'_1} & \cdot & \bar{A}^{(l,l'),(\bar{\sigma}_n)}_{\chi_{n},\chi'_1} \\
         \cdot & \cdot & \cdot & \cdot & \cdot \\
         \cdot & \cdot & \cdot & \cdot & \cdot \\
      \bar{A}^{(l,l'),(\bar{\sigma}_n)}_{\chi,\chi'_n} & \bar{A}^{(l,l'),(\bar{\sigma}_n)}_{\chi_1,\chi'_n} & \bar{A}^{(l,l'),(\bar{\sigma}_n)}_{\chi_2,\chi'_n} & \cdot & \bar{A}^{(l,l'),(\bar{\sigma}_n)}_{\chi_{n},\chi'_n} \\
       \end{array}
     \right).\nonumber\\
\end{eqnarray}
Substituting this equation into Eq.~(\ref{Fll_fermion}) and using the relations
\begin{equation}
\sum^N_{\chi=1}\bar{A}^{(l,l'),(\sigma)}_{\chi' ,\chi}U^{(\bar{\sigma})*}_{\chi l'} =
\begin{cases}
\mathrm{sgn}(l'-l)U^{(\bar{\sigma})*}_{\chi',l'}  &l\neq l',\\
U^{(\bar{\sigma})*}_{\chi',l} & l=l',\\
\end{cases}
	\label{AAU}
\end{equation}
\begin{equation}
\sum^N_{\chi'=1}\bar{A}^{(l,l'),(\sigma)}_{\chi' ,\chi}U^{(\bar{\sigma})}_{\chi' l} =
\begin{cases}
\mathrm{sgn}(l-l')U^{(\bar{\sigma})}_{\chi,l}  &l\neq l',\\
U^{(\bar{\sigma})}_{\chi,l} & l=l',\\
\end{cases}
	\label{AAU1}
\end{equation}
and
\begin{eqnarray}
\sum^N_{\chi=1}\sum^N_{\chi'=1}U^{(\bar{\sigma})*}_{\chi' l'}\bar{A}^{(l,l'),(\sigma)}_{\chi,\chi'}U^{(\bar{\sigma})}_{\chi l}=\delta_{ll'},
\end{eqnarray}
we finally obtain the determinant representation given by Eq.~(\ref{s_ij_1}).

\section{Alternative derivation of Eq.~(\ref{s_ij_1}) using the Cauchy-Binet formula}\label{AppC}
\par In this appendix, we will use the obtained formula of $F^{\mathrm{( p)}}_{j;\vec{\eta}_{n+1},\vec{\chi}_{n}}$ [Eq.~(\ref{SOM_cc_final})] to derive Eq.~(\ref{s_ij_1}). We insert the completeness relation $\sum^{N-1}_{m=0}\sum_{\vec{\eta}_{m+1}}|\vec{\eta}^{\mathrm{(p)}}_{m+1}\rangle\langle \vec{\eta}^{\mathrm{(p)}}_{m+1}|=1$ between $S^-_{l'}$ and $S^+_l$ appearing in $\bar{F}^{\mathrm{( p)}}_{l,l';\vec{\chi}_{n},\vec{\chi}'_{n}}$:
\begin{eqnarray}\label{s_ij_App}
&&\bar{F}^{\mathrm{( p)}}_{l,l';\vec{\chi}_{n},\vec{\chi}'_{n}}\nonumber\\
&=&\sum_{\vec{\eta}_{n+1}}\langle \vec{\chi}^{\mathrm{(p)}}_{n}|S^-_{l'}|\vec{\eta}^{\mathrm{(p)}}_{n+1}\rangle\langle \vec{\eta}^{\mathrm{(p)}}_{n+1}|S^+_{l}|\vec{\chi}^{\mathrm{(p)'}}_{n}\rangle\nonumber\\
&=&\sum_{\vec{\eta}_{n+1}}F^{\mathrm{( p)}*}_{l;\vec{\eta}_{n+1},\vec{\chi}'_{n}}F^{\mathrm{( p)}}_{l';\vec{\eta}_{n+1},\vec{\chi}_{n}}\nonumber\\
&=&\sum_{\vec{\eta}_{n+1}}\det\left(
       \begin{array}{ccccc}
        U^{(\bar{\sigma}_n)*}_{\eta_1,l} & U^{(\bar{\sigma}_n)*}_{\eta_2,l}&\cdot & \cdot & U^{(\bar{\sigma}_n)*}_{\eta_{n+1},l}\\
        A^{(l),(\sigma_n)}_{\eta_1,\chi'_1} & A^{(l),(\sigma_n)}_{\eta_2,\chi'_1} &\cdot & \cdot & A^{(l),(\sigma_n)}_{\eta_{n+1},\chi'_1}\\
        A^{(l),(\sigma_n)}_{\eta_1,\chi'_2} & A^{(l),(\sigma_n)}_{\eta_2,\chi'_2} &\cdot & \cdot & A^{(l),(\sigma_n)}_{\eta_{n+1},\chi'_2}\\
         \cdot & \cdot & \cdot & \cdot & \cdot \\
       A^{(l),(\sigma_n)}_{\eta_1,\chi'_{n}} & A^{(l),(\sigma_n)}_{\eta_2,\chi'_{n}} &\cdot & \cdot & A^{(l),(\sigma_n)}_{\eta_{n+1},\chi'_{n}}\\
       \end{array}
     \right)\nonumber\\
     &&\det\left(
       \begin{array}{ccccc}
        U^{(\bar{\sigma}_n)}_{\eta_1,l'} & A^{(l'),(\sigma_n)*}_{\eta_1,\chi_1} & A^{(l'),(\sigma_n)*}_{\eta_1,\chi_2} & \cdot & A^{(l'),(\sigma_n)*}_{\eta_1,\chi_n} \\
         U^{(\bar{\sigma}_n)}_{\eta_2,l'} & A^{(l'),(\sigma_n)*}_{\eta_2,\chi_1} & A^{(l'),(\sigma_n)*}_{\eta_2,\chi_2} & \cdot & A^{(l'),(\sigma_n)*}_{\eta_2,\chi_n} \\
         \cdot & \cdot & \cdot & \cdot & \cdot \\
         \cdot & \cdot & \cdot & \cdot & \cdot \\
        U^{(\bar{\sigma}_n)}_{\eta_{n+1},l'} & A^{(l'),(\sigma_n)*}_{\eta_{n+1},\chi_1} & A^{(l'),(\sigma_n)*}_{\eta_{n+1},\chi_2} & \cdot & A^{(l'),(\sigma_n)*}_{\eta_{n+1},\chi_n} \\
       \end{array}
     \right).\nonumber\\
\end{eqnarray}
Now we invoke the Cauchy-Binet formula for an $(n+1)\times N$ matrix $\mathcal{A}$ and an $N\times (n+1)$ matrix $\mathcal{B}$:
\begin{eqnarray}\label{Cauchy-Binet}
\det(\mathcal{A}\mathcal{B})=\sum_{\vec{\eta}_{n+1}}\det[\mathcal{A}(\vec{\eta}_{n+1})]\det[\mathcal{B}(\vec{\eta}_{n+1})]
\end{eqnarray}
where $\mathcal{A}(\vec{\eta}_{n+1})$ [$\mathcal{B}(\vec{\eta}_{n+1})$] denote the matrix formed from $\mathcal{A}$ ($\mathcal{B}$) using columns (rows) $(\eta_1,\eta_2,\cdots,\eta_{n+1})$. We see that the two matrices
\begin{eqnarray}\label{Cauchy-Binet_A}
\mathcal{A}&=&  \left(
       \begin{array}{ccccc}
        U^{(\bar{\sigma}_n)*}_{1,l} & U^{(\bar{\sigma}_n)*}_{2,l}&\cdot & \cdot & U^{(\bar{\sigma}_n)*}_{{N},l}\\
        A^{(l),(\sigma_n)}_{1,\chi'_1} & A^{(l),(\sigma_n)}_{2,\chi'_1} &\cdot & \cdot & A^{(l),(\sigma_n)}_{{N},\chi'_1}\\
        A^{(l),(\sigma_n)}_{1,\chi'_2} & A^{(l),(\sigma_n)}_{2,\chi'_2} &\cdot & \cdot & A^{(l),(\sigma_n)}_{{N},\chi'_2}\\
         \cdot & \cdot & \cdot & \cdot & \cdot \\
       A^{(l),(\sigma_n)}_{1,\chi'_{n}} & A^{(l),(\sigma_n)}_{2,\chi'_{n}} &\cdot & \cdot & A^{(l),(\sigma_n)}_{{N},\chi'_{n}}\\
       \end{array}
     \right)
\end{eqnarray}
and
\begin{eqnarray}\label{Cauchy-Binet_B}
\mathcal{B}&=& \left(
       \begin{array}{ccccc}
        U^{(\bar{\sigma}_n)}_{1,l'} & A^{(l'),(\sigma_n)*}_{1,\chi_1} & A^{(l'),(\sigma_n)*}_{1,\chi_2} & \cdot & A^{(l'),(\sigma_n)*}_{1,\chi_n} \\
         U^{(\bar{\sigma}_n)}_{2,l'} & A^{(l'),(\sigma_n)*}_{2,\chi_1} & A^{(l'),(\sigma_n)*}_{2,\chi_2} & \cdot & A^{(l'),(\sigma_n)*}_{2,\chi_n} \\
         \cdot & \cdot & \cdot & \cdot & \cdot \\
         \cdot & \cdot & \cdot & \cdot & \cdot \\
        U^{(\bar{\sigma}_n)}_{{N},l'} & A^{(l'),(\sigma_n)*}_{{N},\chi_1} & A^{(l'),(\sigma_n)*}_{{N},\chi_2} & \cdot & A^{(l'),(\sigma_n)*}_{{N},\chi_n} \\
       \end{array}
     \right)\nonumber\\
\end{eqnarray}
fit into Eq.~(\ref{s_ij_App}) and gives
\begin{eqnarray}
&&\bar{F}^{\mathrm{( p)}}_{l,l';\vec{\chi}_{n},\vec{\chi}'_{n}}=\det(\mathcal{A}\mathcal{B}),
\end{eqnarray}
which is exactly Eq.~(\ref{s_ij_1}) after using Eq.~(\ref{AU_U}).
\section{Proof of $\bar{F}^{\mathrm{( p)}}_{l,l';\vec{j}_{n},\vec{j}'_{n}}=\langle \vec{j}_n|S^+_{l}S^-_{l'}|\vec{j}'_n\rangle$}\label{AppD}
We only consider $l< l'$ as the case of $l>l'$ can be proved similarly. In the atomic limit, the eigenbasis is consistent with the position basis spanned by the Ising configurations, i.e., $\vec{\chi}_n=\vec{j}_n$ and $\vec{\chi}'_n=\vec{j}'_n$. From Eq.~(\ref{sum_AA}), we have $\bar{A}^{(j,j'),(\sigma)}_{\eta,\eta'} =\delta_{\eta,\eta'}(1-2 \sum^{j_{\max}-1}_{l=j_{\min}} \delta_{\eta',l})$. After performing a Laplace expansion of the determinant in Eq.~(\ref{s_ij_1}) along the first row, we arrive at
\begin{eqnarray}\label{Flljj}
&&\bar{F}^{\mathrm{( p)}}_{l,l';\vec{j}_{n},\vec{j}'_{n}}= \sum^n_{m=1} \sum^n_{m'=1}(-1)^{m+m'} \delta_{j_m,l} \delta_{j'_{m'},l'}   \delta_{\vec{j}^{(m)}_n,\vec{j}'^{(m')}_n}\nonumber\\
&&\prod^n_{m''=1,(\neq m')}\left(1-2 \sum^{l'-1}_{l''=l} \delta_{j'_{m''},l''}\right).
\end{eqnarray}
We first observe that the product $\prod^n_{m''=1,(\neq m')}$ in the last line of the above equation can be replaced by $\prod^{m'-1}_{m''=1}$ because $j'_{m''}>j'_{m'}=l'$ for $m''\geq m'+1$. We then argue that only terms with $m\leq m'$ contribute to the double summation in Eq.~(\ref{Flljj}). Actually, if $m>m'$, then from $\vec{j}^{(m)}_n=\vec{j}'^{(m')}_n$ we have $j_{m'-1}=j'_{m'-1}$ and $j_{m'}=j'_{m'+1}$, so that $l'=j'_{m'}<j'_{m'+1}=j_{m'}<j_m=l$, in contradiction to the assumption that $l<l'$. Thus, for $l<l'$ we have
\begin{eqnarray}\label{Flljj1}
&&\bar{F}^{\mathrm{( p)}}_{l,l';\vec{j}_{n},\vec{j}'_{n}}= \sum^n_{m\leq m'}(-1)^{m+m'} \delta_{j_m,l} \delta_{j'_{m'},l'}   \delta_{\vec{j}^{(m)}_n,\vec{j}'^{(m')}_n}\nonumber\\
&&\prod^{m'-1}_{m''=1}\left(1-2 \sum^{l'-1}_{l''=l} \delta_{j'_{m''},l''}\right).
\end{eqnarray}
For $m=m'$, we have $j_{m''}=j'_{m''}$ for $m''\leq m'-1$, so that $j'_{m''}=j_{m''}<j_{m}=l$, and hence $\prod^{m'-1}_{m''=1}\left(1-2 \sum^{l'-1}_{l''=l} \delta_{j'_{m''},l''}\right)=1$; while for $m<m'$, we have $j_{m''}=j'_{m''}$ for $m''\leq m-1$, and $j_{m+1}=j'_m ,~\cdots, j_{m'}=j'_{m'-1}$, so that $l=j_m<j_{m+1}=j'_m<\cdots\leq j'_{m'-1}\leq l'-1$, and hence $\prod^{m'-1}_{m''=1}\left(1-2 \sum^{l'-1}_{l''=l} \delta_{j'_{m''},l''}\right)=\prod^{m'-1}_{m''=m}\left(1-2 \sum^{l'-1}_{l''=l} \delta_{j'_{m''},l''}\right)=(-1)^{m'-m}$. We therefore always have
\begin{eqnarray}\label{Flljj2}
&&\bar{F}^{\mathrm{( p)}}_{l,l';\vec{j}_{n},\vec{j}'_{n}}= \sum^n_{m\leq m'}\delta_{j_m,l} \delta_{j'_{m'},l'}   \delta_{\vec{j}^{(m)}_n,\vec{j}'^{(m')}_n},
\end{eqnarray}
which states that for two Ising configurations $|\vec{j}_n\rangle$ and $|\vec{j}'_n\rangle$ that satisfy $j_{m}\notin\{\vec{j}'_n\}$ and $j'_{m}\notin\{\vec{j}_n\}$, as well as $\vec{j}^{(m)}_n=\vec{j}'^{(m')}_n$, the matrix element $\bar{F}^{\mathrm{( p)}}_{l,l';\vec{j}_{n},\vec{j}'_{n}}$ is just unity, consistent with the definition $\bar{F}^{\mathrm{( p)}}_{l,l';\vec{j}_{n},\vec{j}'_{n}}=\langle \vec{j}_n|S^+_{l}S^-_{l'}|\vec{j}'_n\rangle$.


\begin{thebibliography}{99}
\bibitem{Bethe} H. Bethe, Z. Physik \textbf{71}, 205 (1931).
\bibitem{LSM} E. Lieb, T. Schultz, and D. Mattis, Ann. Phys. (NY) \textbf{16}, 407 (1961).
\bibitem{2DIsing} T. D. Schultz, and D. C. Mattis, and E. H. Lieb, Rev. Mod. Phys. \textbf{36}, 856 (1964).
\bibitem{Iorgov} N. Iorgov, V. Shadura, Yu. Tykhyy, J. Stat. Mech., \textbf{2011}, P02028 (2011).
\bibitem{BHM} J. Jordan, R. Orus, and G. Vidal, Phys. Rev. B \textbf{79}, 174515 (2009).
\bibitem{Stolze} P. Karbach and J. Stolze, Phys. Rev. A \textbf{72}, 030301(R) (2005).
\bibitem{fleming} Y.-C. Cheng and G. R. Fleming, Annu. Rev. Phys. Chem. \textbf{60}, 241 (2009).
\bibitem{Rydberg} D. Barredo, H. Labuhn, S. Ravets, T Lahaye, A. Browaeys, and C. S. Adams, Phys. Rev. Lett. \textbf{114}, 113002 (2015).
\bibitem{Spano1991} F. C. Spano, Phys. Rev. Lett. \textbf{67}, 3424 (1991).
\bibitem{Hanamura} T. Tokihiro, Y. Manabe, and E. Hanamura, Phys. Rev. B \textbf{47}, 2019 (1993).
\bibitem{PRA2014} N. Wu, A. Nanduri, and H. Rabitz, Phys. Rev. A \textbf{89}, 062105 (2014).
\bibitem{PRB2016} N. Wu, N. Fr\"ohling, X. Xing, J. Hackmann, A. Nanduri, F. B. Anders and H. Rabitz, Phys. Rev. B \textbf{93}, 035430 (2016).
\bibitem{ratchet2015} K. D. B. Higgins, B. W. Lovett, and E. M. Gauger, J. Phys. Chem. C \textbf{121}, 20714 (2017).
\bibitem{math} https://math.stackexchange.com/questions/1286193/how-to-prove-the-following-determinant-identity
\bibitem{2016LPP} N. Wu, J. Feist, and F. J. Garc\'ia-Vidal, Phys. Rev. B \textbf{94}, 195409 (2016).
\bibitem{Loss2009} O. Tsyplyatyev and D. Loss, Phys. Rev. A \textbf{80}, 023803 (2009).
\bibitem{Loss2010} O. Tsyplyatyev and D. Loss, Phys. Rev. B \textbf{82}, 024305 (2010).
\bibitem{PRA2009} K. H\"ark\"onen, F. Plastina, and S. Maniscalco, Phys. Rev. A \textbf{80}, 033841 (2009).
\bibitem{Fari2012} C. Str\"ater, O. Tsyplyatyev, and A. Faribault, Phys. Rev. B \textbf{86}, 195101 (2012).
\bibitem{Dicke} R. H. Dicke, Phys. Rev. \textbf{93}, 99 (1954).
\bibitem{fnote} We note that there was a typo in Eq.~(4.8b) of Ref.~\cite{Hanamura}: the factor $\prod_{j>j'}(e^{-ik'_j}-e^{ik'_{j'}})$ in the numerator of Eq.~(4.8b) should be $\prod_{j>j'}(e^{-ik'_j}-e^{-ik'_{j'}})$.
\bibitem{Hanamura1995} T. Tokihiro, Y. Manabe, and E. Hanamura, Phys. Rev. B \textbf{51}, 7655 (1995).
\bibitem{mu12} We note that this result is inconsistent with Ref.~\cite{Spano1991}, where it was claimed that $\mu^{\mathrm{(o)}}_{\eta,\chi_1\chi_2}$ is nonzero only if $\chi_1+\chi_1$ is odd.
\bibitem{Spano-Mu} F. C. Spano and S. Mukamel, Phys. Rev. Lett. \textbf{66}, 1197 (1991).
\bibitem{fifth} A. A. Said, C. Wamsley, D. J. Hagan, E. W. Van Stryland, B. A. Reinbardt, P. Roderer, and A. G. Dillard, Chemical Physics Letters \textbf{228}, 646 (1994).
\bibitem{NPB} W. V. Pogosov, D. S. Shapiro, L. V. Bork, and A. I. Onishchenko, Nuclear Physics B \textbf{919}, 218 (2017).
\bibitem{Spanoreview} F. C. Spano, Acc. Chem. Res. \textbf{43}, 429 (2010).
\end{thebibliography}
\end{document}